\documentclass[12pt]{article} 
\usepackage[utf8]{inputenc}
\usepackage{enumerate}
\usepackage{ytableau}
\usepackage{amsfonts}
\usepackage{tikz}
\usepackage{blkarray}  
\usepackage[hyperfootnotes=false]{hyperref}
\usepackage{amsmath}
\usepackage{amssymb}
\usepackage{graphicx}
\usepackage{bm}
\usepackage{cite}
\usepackage{braket}
\usepackage{wasysym}
\usepackage{mciteplus} 
\usepackage{subfigure}
\usepackage{subcaption}
\usepackage{bbold}
\usepackage{slashed,extarrows}

\renewcommand{\title}[1]{\vbox{\center\bf{\Large{#1}}}\vspace{5mm}}
\renewcommand{\author}[1]{\vbox{\center#1}\vspace{5mm}}

\newlength{\abstractwidth}
\DeclareMathOperator{\sech}{sech}
\DeclareMathOperator{\arccosh}{arccosh}
\DeclareMathOperator{\csch}{csch}

\begin{document}
	
	\def\im{\text{i}} \def\eqa{\begin{eqnarray}}
		\def\eqae{\end{eqnarray}} \def\be{\begin{equation}}
		\def\ee{\end{equation}} \def\bea{\begin{eqnarray}}
		\def\eea{\end{eqnarray}}\def\ba{\begin{array}}
		\def\ea{\end{array}}\def\bd{\begin{displaymath}}
		\def\ed{\end{displaymath}}\def\bal#1\eal{\begin{align}#1\end{align}}
	\def\eg{{\it e.g.~}}
	\def\ie{{\it i.e.~}}
	\def\Tr{{\rm Tr}}
	\def\tr{{\rm tr}}
	\def\>{\rangle}
	\def\<{\langle}
	\def\a{\alpha}
	\def\b{\beta}
	\def\c{\chi}
	\def\del{\delta}
	\def\e{\epsilon}
	\def\f{\phi}
	\def\vf{\varphi}
	\def\tvf{\tilde{\varphi}}
	\def\g{\gamma}
	\def\h{\eta}
	\def\j{\psi}
	\def\k{\kappa}
	\def\l{\lambda}
	\def\m{\mu}
	\def\n{\nu}
	\def\w{\omega}
	\def\p{\pi}
	\def\q{\theta}
	\def\r{\rho}
	\def\s{\sigma}
	\def\t{\tau}
	\def\u{\upsilon}
	\def\x{\xi}
	\def\z{\zeta}
	\def\D{\Delta}
	\def\F{\Phi}
	\def\G{\Gamma}
	\def\J{\Psi}
	\def\L{\Lambda}
	\def\W{\Omega}
	\def\P{\Pi}
	\def\Q{\Theta}
	\def\S{\Sigma}
	\def\U{\Upsilon}
	\def\X{\Xi}
	\def\nab{\nabla}
	\def\pa{\partial}
	\def\calT{\mathcal{T}}
	\def\cM{\mathcal{M}}
	\def\pcM{\partial\mathcal{M}}
	\def\({\left(}
	\def\){\right)}
	\def\[{\left[}
	\def\]{\right]}
	\def\nn{\nonumber \\}
	\def\d{\operatorname{d}}
	\def\ttbar{T$\overline{\text{T}}$ }
	\def\Renyi{R$\acute{\text{e}}$nyi }
	\def\Poincare{Poincar$\acute{\text{e}}$ }
	\def\Banados{Ba$\tilde{\text{n}}$ados }
	\def\arctan{\operatorname{arctan}}
	\def\arctanh{\operatorname{arctanh}}
	\def\wa{\rm I^{W}_{AdS}}
	
	
	\newcommand{\la}{\leftarrow}
	\newcommand{\ra}{\rightarrow}
	\newcommand{\lra}{\leftrightarrow}
	\newcommand{\bc}{{\mathbb{C}}}
	\newcommand{\br}{{\mathbb{R}}}
	\newcommand{\bz}{{\mathbb{Z}}}
	\newcommand{\bp}{{\mathbb{P}}}
	\newcommand{\OL}[1]{ \hspace{1pt}\overline{\hspace{-1pt}#1
			\hspace{-1pt}}\hspace{1pt} }
	\newcommand{\OO}{\mathcal{O}}

	\newcommand{\red}{\textcolor[RGB]{255,0,0}}
	\newcommand{\blue}{\textcolor[RGB]{0,0,255}}
	\newcommand{\green}{\textcolor[RGB]{0,255,0}}
	\newcommand{\cyan}{\textcolor[RGB]{0,255,255}}
	\newcommand{\magenta}{\textcolor[RGB]{255,0,255}}
	\newcommand{\yellow}{\textcolor[RGB]{255,255,0}}
	\newcommand{\sky}{\textcolor[RGB]{135, 206, 235}}
	\newcommand{\orange}{\textcolor[RGB]{255, 127, 0}}
	\definecolor{darkgreen}{rgb}{0,.5,0}
	\newcommand{\DR}[1]{{\sl{\textcolor{blue}{#1}}}}
	\newcommand{\DS}[1]{{\sl{\textcolor{darkgreen}{#1}}}}
	
\begin{titlepage}
\bigskip

\bigskip\bigskip

\bigskip
			
\begin{center}
\centerline
{\Large \bf {Holographic correlation functions from wedge}}
\bigskip
				
\bigskip
				
\end{center}
			
\begin{center}
\bf{{Tengzhou Lai$^{a,}$}\footnote{laitengzhou20@mails.ucas.ac.cn}, {Ya-Wen Sun$^{a, c,}$}\footnote{yawen.sun@ucas.ac.cn}, {Jia Tian$^{b, c,}$}\footnote{wukongjiaozi@ucas.ac.cn}}\\
\bigskip

%
				
\bigskip
\bigskip 
{\it $^a$School of Physical Sciences, University of Chinese Academy of Sciences, \\Zhongguancun East Road 80, Beijing 100190, P.~R.~China\\
	\vspace{2mm}
$^b$State Key Laboratory of Quantum Optics and Quantum Optics Devices, Institute of Theoretical Physics, Shanxi University, Taiyuan 030006, P.~R.~China
\\
\vspace{2mm}
$^c$Kavli Institute for Theoretical Sciences (KITS), University of Chinese Academy \\ of Science, 100190 Beijing, P.~R.~China 
}
\vspace{10mm}
\end{center}

\bigskip\bigskip
\begin{abstract}
In this work, we propose a novel holographic method for computing correlation functions of operators in conformal field theories. This method refines previous approaches and is specifically aimed at being applied to heavy operators. For operators that correspond to particles in the bulk, we show that the correlation functions can be derived from the on-shell actions of excised geometries for heavy operators, using numerical and perturbative calculations. These excised geometries are constructed from various background solutions such as \Poincare AdS$_3$, global AdS$_3$, and BTZ by cutting out a wedge bounded by two intersecting End-of-the-world branes and the AdS boundary. The wedge itself can be interpreted as a dual to a BCFT with cusps in the AdS/BCFT framework. Additionally, we calculate the correlation functions for heavy operators directly by constructing backreacted bulk geometries for particle excitations through coordinate transformations from a conical solution. We find that the on-shell actions of these backreacted solutions accurately reproduce correlation functions, although they differ from those computed in Fefferman-Graham(FG) gauge. This discrepancy, previously noted and explained in our earlier work, is reinforced by additional examples presented here.

\medskip
\noindent
\end{abstract}
\bigskip \bigskip \bigskip
			
\vspace{1cm}
			
\vspace{2cm}
			
\end{titlepage}
\tableofcontents
		
\section{Introduction} 
\label{sec:intro}
\renewcommand{\theequation}{1.\arabic{equation}}
\setcounter{equation}{0}
Over the past few decades, string theory has been the most popular candidate for a quantum theory of gravity. One of its most significant implications is the AdS/CFT correspondence \cite{Maldacena:1997re,Aharony:1999ti,Witten:1998qj}, which states that  a quantum gravity theory living in the bulk AdS$_{d+1}$ space is dual to a conformal field theory CFT$_{d}$ defined on the asymptotic AdS boundary. Various aspects of this correspondence have been validated \cite{Strominger:1996sh,Klebanov:1999tb,Susskind:1998dq,Banks:1998dd,Horowitz:1999gf}, with a crucial task being the understanding of the holographic dual of correlation functions, which are fundamental observables in conformal field theories. The standard method for computing correlation functions is encapsulated in the GKPW dictionary \cite{Witten:1998qj,Gubser:1998bc}:
\bea
Z_{\rm AdS}\(\f_{\pa}=\f_0\)=\left<\text{exp}\(\int \f_0\mathcal{O}\)\right>_{\rm CFT}
\eea
where the boundary value of a bulk field provides the source for the dual operator inserted on the boundary CFT. However, an implicit assumption in this framework is that the ADM mass of the bulk field, or the scaling dimension of the dual primary operator, remains of constant order, specifically $\Delta \sim \mathcal{O}(1) \ll \frac{\ell^{d-1}}{G_N}$. This condition implies that the backreaction from the bulk fields is negligible, allowing for reliable calculations \cite{Freedman:1998tz} and conclusions based on a fixed background.  When the scaling dimension of the dual operator increases but remains bounded such that $\mathcal{O}(1) \ll \Delta \ll \frac{\ell^{d-1}}{G_N}$, the corresponding correlation function can be approximated using a classical saddle point or WKB approximation \cite{Balasubramanian:1999zv, Balasubramanian:2012tu,Kastikainen:2021ybu}:
\be
\left\langle\mathcal{O}(x_1)\mathcal{O}(x_2)\right\rangle=\int_{\mathcal{P}}D[\mathcal{P}]e^{-\D L(\mathcal{P})}\sim \sum_{\rm geodesics} e^{-\D L(x_1,x_2)}
\ee
where $\mathcal{P}$ represents arbitrary bulk path connecting two insertions and  $L(\mathcal{P})$ is the corresponding regularized length functional. Once the scaling dimension of the dual operator approaches the CFT central charge, i.e., $\Delta \sim \frac{\ell^{d-1}}{G_N}$, the backreaction on the bulk geometry becomes significant, and the on-shell action encodes all pertinent information about the correlation functions or a specific conformal block \cite{Chang:2016ftb, Chen:2016dfb, Alkalaev:2017bzx, Hijano:2015qja, Hijano:2015zsa}. Recently, a spacetime banana proposal was introduced in \cite{Abajian:2023jye,Abajian:2023bqv}, refined and generalized in \cite{Tian:2024fmo}, for computing correlation functions of so-called huge operators that are sufficiently heavy to deform the geometry into a black hole.

However, usually constructing backreacted geometries is a very challenging task. A relatively simple situation is when the inserted operators are dual to massive particles, where the backreacted geometries are described by conical solutions. Nevertheless, for the purpose of calculating correlation functions, we need to perform a coordinate transformation to transform the geometry to a form in accordance with the calculation of the two-point function, with the trajectory of the particle excitation connecting the two spacetime points where the boundary operators are located. The explicit metric after this transformation is still very hard to find, especially when we calculate higher points correlation functions. One resolution is based on the fact that the solution near the boundary is fixed by its behaviors around each boundary insertion point. As a result, its on-shell action can be directly computed with no need of the details of the whole metric \cite{Krasnov:2000zq,Hadasz:2005gk} and it turns out the action reduces to the one of a Liouville theory \cite{Harlow:2011ny,Abajian:2023bqv}. 

In this paper, to resolve this problem we propose a more direct approach to compute the on-shell actions of conical solutions by leveraging the fact that these geometries can be constructed through a cut-and-glue procedure \cite{Balasubramanian:1999zv,Matschull:1998rv,Louko:2000tp}. We demonstrate that the correlation functions could be read from the on-shell action of an excised geometry, thereby confirming the cut-and-glue construction at the level of matching on-shell actions. More specifically, our proposal is quantitatively expressed as \footnote{The meaning of each term will be clarified in the next section.} 
\be
I^{BR}=I^{AdS_3}-I_{gravity,W}+I_{m}
\ee
where the left-hand side represents the on-shell action of a general backreacted conical geometry with bulk particle excitations, and the right-hand side corresponds to the excised geometry. It is important to note that this equivalence has been examined in \cite{Caputa:2022zsr} for a specific example, and a two-point correlation function could be read from the on-shell action in this way. In this work, based on their discoveries, we formulate this wedge proposal explicitly, and utilize this equivalence at the level of the on-shell action to derive the two-point correlation function of dual heavy operators. The correct form of correlations functions obtained further confirm this equivalence and the correctness of this wedge proposal, with the hope of generalizing it to higher-point correlation function in future studies.

We will explore this wedge proposal across various AdS$_3$ solutions, highlighting caveats and subtleties. Though the wedge proposal could, in principle, be generalized to higher dimensions, the advantage of focusing on AdS$_3$ solutions lies in the fact that conical geometries associated with bulk particle excitations can be obtained through coordinate transformations in three dimensions, as demonstrated in \cite{Nozaki:2013wia}. Therefore, we can verify the wedge proposal through this direct calculation, motivating us to also directly compute the two-point functions in the backreacted geometries. In our previous study \cite{Tian:2024fmo} on the on-shell action for the backreacted \Poincare geometry,  an unexpected result was uncovered: the on-shell actions of the backreacted geometry in transformed coordinates differ from those in FG \cite{graham1985charles,Fefferman:2007rka} coordinates. Here in all direct calculations of the two-point functions in the backreacted geometries, we will show that this discrepancy is common and manifests in all examined examples.

The outline of this paper is listed as follows: in section \ref{sec:wedge_proposal}, we give our wedge proposal of correlation functions and relate it to the AdS/BCFT formalism \cite{ Takayanagi:2011zk, Fujita:2011fp}; then in section \ref{sec:cases}, we analytically compute the on-shell action inside a wedge region case by case in various AdS$_3$ spacetime both analytically and numerically, and confirm their consistence. The result obtained successfully reproduces the two-point correlation function. In section \ref{sec:backreacted geometries}, we straightforwardly construct backreacted geometries and compute the on-shell action both in the whole coordinate patch and the FG patch. In section \ref{sec:discussion}, we briefly discuss our result and give a few future directions.

\section{The Wedge proposal of correlation functions}
\label{sec:wedge_proposal}
\renewcommand{\theequation}{2.\arabic{equation}}
\setcounter{equation}{0}
In this section, we propose a novel holographic approach to compute the correlation functions of operators in conformal field theories (CFTs). Our method integrates two key insights recently revisited in \cite{Caputa:2022zsr} and \cite{Abajian:2023bqv,Abajian:2023jye,Tian:2024fmo}. Firstly, the correlation functions or conformal blocks can be derived from the on-shell action of gravity on specific conical geometries for heavy operators corresponding to particle excitations in the bulk. For instance, as discussed in \cite{Abajian:2023bqv,Abajian:2023jye,Tian:2024fmo}, the holographic dual of the two-point function can be constructed from the standard conical AdS solution by applying a global to \Poincare (GtP) transformation\cite{Abajian:2023bqv}
\be
r=\frac{R}{z},\quad \tau=\frac{1}{2}\log\(R^2+z^2\)
\ee
followed by a special conformal transformation(SCT). Secondly, it is well established that a conical geometry can be interpreted as a wedge of the global AdS geometry or equivalently as the global AdS geometry with a wedge region excised \cite{Balasubramanian:1999zv,Matschull:1998rv,Louko:2000tp}.

Combining these two insights suggests that the correlation functions may also be calculated from the on-shell action of a bulk geometry with a specific wedge excised. This idea, to our knowledge, was first articulated in \cite{Caputa:2014eta}, albeit with some caveats, which we will review below. In this paper, we aim to explore more general scenarios. A subtlety in the wedge construction of conical geometry is the lack of rigorous proof that the equivalence between the excised geometry and the conical geometry holds at the level of the on-shell action. It turns out that to assert this equality, additional terms must be added to the action of the gravity theory. For instance, to accurately describe a spacetime wedge, it is more appropriate to employ the AdS/BCFT formalism, as the wedge possesses spacetime boundaries. Furthermore, the tip of the wedge represents another type of spacetime singularity, which also necessitates consideration in the action.

The advantage of our wedge approach is that it circumvents the need to construct the specific backreacted geometry for the calculation of the correlation functions with operator insertions, which is often a challenging task. For example, we can conveniently utilize the wedge construction to describe the formation of a black hole through the collision of two particles \cite{Matschull:1998rv}.
		
\subsection{A toy model}
To illustrate these issues, let us consider a toy model: the conical AdS which has the metric
\be\label{conical AdS3}
ds^2=\(r^2+\a^2\)d\tau^2+\frac{dr^2}{r^2+\a^2}+r^2d\theta^2,\quad \q\sim\q+2\pi.
\ee
The geometry is not smooth but has a conical singularity at $r=0$ with a deficit angle $2\pi (1-\alpha)$, which can be thought of as the backreaction of a  massive particle that is dual to a heavy operator in the boundary field theory, so the total action of the system is 
\be\label{action_gravity}
I_{\text{grav}}=-\frac{1}{16\pi G_N}\int_{\cal{M}}\sqrt{g}\(R+2\)-\frac{1}{8\pi G_N}\int_{\pa\cal{M}}\sqrt{h}\(K-1\)+m\int_{\Gamma}\sqrt{\gamma}\,,
\ee
where $\Gamma$ represents the trajectory of the particle and $m$ is the local mass of the particle. In particular, the profile $\G$ is stationary and along with the AdS center in this case, the mass and the deficit angle is related by the component of Einstein equation \cite{Fursaev:1995ef} along the trajectory
\be
R_{\t\t}-\frac{1}{2}g_{\tau\tau}R-g_{\tau\tau}=-2\pi\(1-\a\)\delta_{\Gamma}g_{\tau\tau}
\ee
which implies that 
\be\label{relation}
m=\frac{1-\a}{4G_N}\, .
\ee
Therefore, the on-shell action equals
\bea 
I_{\text{grav}}&=&-\frac{1}{16\pi G_N}\int_{\cM/\Gamma}\sqrt{g}(R+2)-\frac{1}{8\pi G_N}\int_{\pcM} \sqrt{h}(K-1),\\
&=&\frac{1}{4\pi G_N}\int_{\pcM}\(\frac{\Lambda^2-\epsilon^2}{2}\)-\frac{1}{8\pi G_N}\int_{\pcM}\(\Lambda^2+\frac{\a^2}{2}\)\\
&=&-\frac{\a^2}{16\pi G_N} W T=-\frac{\a^2}{8 G_N} \xi , \label{on1}
\eea 
where $W=2\pi$ is the width of the boundary cylinder, $T$ is the length of the boundary cylinder and $\xi=2\pi T/W$ is the dimensionless modular parameter. From the on-shell action, we read off the excitation energy due to the particle insertion is $E=(-\frac{\alpha^2}{8G_N})-E_0=-\frac{c}{12}\alpha^2+\frac{c}{12})=\frac{c}{12}(1-\alpha^2)$, where the relation $\frac{1}{2G_N}=\frac{c}{3}$ \cite{Brown:1986nw} and the vacuum energy is $-\frac{c}{12}$ have been used.

\bigskip
		
Another description of the backreaction on the geometry is to excise a bulk wedge with angle $2\pi(1- \a)$ \cite{Matschull:1998rv,Balasubramanian:1999zv} from the global AdS geometry. To describe the wedge, we consider the \bea \label{BCFTaction}
I_{\text{gravity},W}&=&-\frac{1}{16\pi G_N}\int_\cM \sqrt{g}(R+2)-\frac{1}{8\pi G_N}\int_{\pcM}\sqrt{h}(K-1)-\frac{1}{8\pi G_N}\int_{Q_1\cup Q_2}\sqrt{h}(K-T)\nn
&&-\frac{1}{8\pi G_N}\int_\Gamma\sqrt{\gamma}\cos^{-1}(n_{1}\cdot n_2)-\frac{1}{8\pi G_N}\int_{Q_{1,2}\cap \pcM}\sqrt{\gamma}\cos^{-1}\(n_{1,2}\cdot n_{0}\)
\eea
where $n_{1,2}$ denotes the normal vectors to the End of the world(EOW) branes $Q_{1,2}$ ($\theta=\pm\pi(1-\a)$) with tension $T=0$, while $n_0$ represents the normal vector to the asymptotic boundary. As a result, the boundaries $Q_{1,2}$ of wedge  do not contribute to the action. Additionally, since the EOW branes are perpendicular to the AdS boundary, the Hayward terms \cite{Hayward:1993my} at $Q_{1,2}\cap\pcM$ do not contribute either. The only non-trivial term is the corner term  at the tip of the wedge\footnote{Here, due to the identification of these two edges, we should subtract an overall angle of $\pi$ as in \cite{Caputa:2014eta,Boruch:2021hqs}.}
\bea 
I_{\text{Hayward}}&=&-\frac{1}{8\pi G_N}\int_\Gamma\sqrt{\gamma}\cos^{-1}(n_{1}\cdot n_2)\nn
&=&-\frac{1}{8\pi G_N}\int_\Gamma \sqrt{\gamma}\(\pi-2\pi(1-\a)-\pi\)=\frac{1-\alpha}{4G_N}\int_\Gamma\sqrt{\gamma},
\eea 
which precisely cancels the mass term associated with the particles. Therefore, the on-shell action is
\bea
I^{AdS_3}-I_{gravity,W}+I_{m}=-\frac{1}{16\pi G_N}W^{}T=-\frac{\alpha^2}{8 G_N}\xi, \label{on2}
\eea
where the modular parameter $\xi=2\pi T/W$ and $W=2\pi \alpha$. Thus, comparing \eqref{on1} with \eqref{on2} reveals that these two descriptions agree with each other on the level of on-shell action. The important lesson is that in both descriptions, the mass term always gets canceled by the term that describes the conical singularity \cite{Chandra:2022bqq}. It means when we compute the on-shell action, we do not consider them \footnote{For convenience, we will use $I_{gravity,W}$ to represent $I_{gravity,W}-I_{m}$ in the following context.}. Before delving into the wedge constructions, let us first introduce the theory for properly describing the wedge.
		
\subsection{AdS/BCFT with corners}
The proper theory for describing a wedge is the generalized AdS/BCFT duality in which the EOW branes are connected or intersected along a defect. The action of the gravity model is given by
\bea \label{gravity}
I=-\frac{1}{16\pi G_N}\int_{\cM} \sqrt{g}(R+2)-\frac{1}{8\pi G_N}\int_{\text{EOW}}\sqrt{h}(K-T)+\frac{1}{8\pi G_N}\int_\Gamma(\theta-\theta_0)\sqrt{\gamma},
\eea 
where the last term is the analog of the Hayward term  and $\theta_0$ is a fixed value that characterizes the corner. This generalization was introduced in \cite{Geng:2021iyq,Kusuki:2022ozk,Miyaji:2022dna,Biswas:2022xfw}, and further explored in \cite{Tian:2023vbi} (also see our companion paper \cite{Tian:2024fmo} for a novel application). With this generalization, we can construct a holographic dual of a boundary-condition-changing (BCC) operator \cite{cardy1989boundary,Cardy:2004hm}, defined as the primary state with the lowest conformal dimension in a BCFT defined on an interval with a pair of different boundary conditions. In the standard AdS/BCFT framework, the gravitational description of a BCFT on an interval $\theta\in[0,\pi]$ in the low-temperature phase is the global AdS$_3$ solution with the metric 
\be\label{global_metric}
ds^2=(r^2+1)d\tau^2+\frac{dr^2}{r^2+1}+r^2d\theta^2,\quad \theta\sim \theta+2\pi.
\ee 
The two boundaries at $\theta=0,\pi$ correspond to a EOW brane with the profile function
\be
r\sin\q =-\l, \quad \l\equiv \frac{T}{\sqrt{1-T^2}}
\ee 
where $T$ is the tension of the brane. Obviously, the EOW brane intersects the AdS boundary at the $\theta=0$ and $\theta=\pi$ as desired. Because the two boundaries are connected by the same EOW brane, effectively, we are imposing the same conformal boundary condition on them. Therefore, the BCC operator in this case is just the identity operator. To verify this, we can compute the on-shell action \eqref{gravity} of this bulk solution, the result is 
\be 
I=-\frac{\beta }{16 G_N}=-\frac{c}{24}\beta,
\ee 
where $\beta$ is the size of the $\tau$-circle. Recall that the vacuum energy of is $E_0=-\frac{c}{24}$, we conclude that the conformal dimension of the BCC operator is $-\frac{c}{24}-E_0=0$, indicating that the BCC operator here is the identity operator.  
		
\begin{figure}[htbp]
	\begin{center}
	 \includegraphics[scale=0.45]{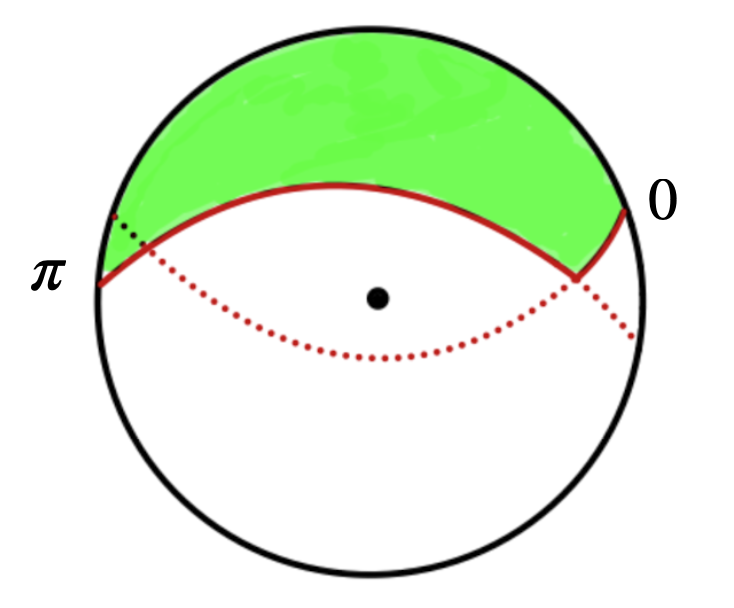}
	\caption{The bulk dual of a non-trivial BCC operator.}\label{bcc}
	\end{center}
\end{figure}

\begin{figure}[h]
	\centering
\includegraphics[width=0.55\textwidth]{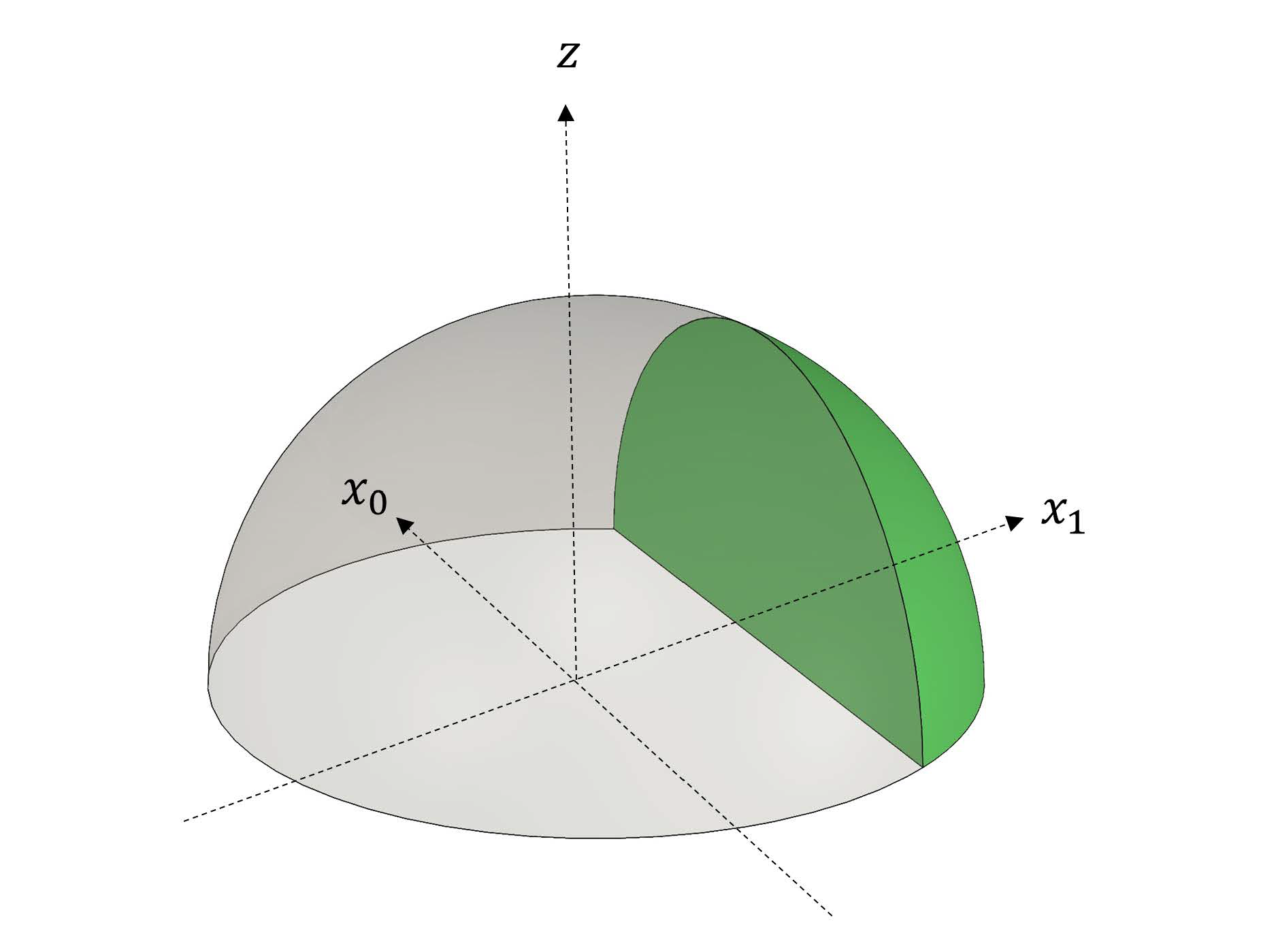}
\caption{A simple gravity solution dual to the two-point correlation function of BCC operators. The shaded green region represents the bulk dual, bounded by two EOW branes and the AdS boundary.}
\label{EOW1}
\end{figure}

To find the gravity dual of a non-trivial BCC operator, as shown in \cite{Miyaji:2022dna,Biswas:2022xfw}, we should consider the conical AdS$_3$ bulk solution with the metric \eqref{conical AdS3}. The EOW brane in the conical AdS geometry has the general profile
\be 
r\sin[\alpha(\theta-\theta_0)]=-\a\l,\quad \theta\in(\theta_0,\theta_0+\frac{\pi}{\alpha}).
\ee 
Therefore, requiring that the two boundaries of the BCFT are still at $\theta=0,\pi$, we have to use two EOW branes. This is illustrated in Figure \ref{bcc}. Without the loss of generality, we can choose the two EOW branes to be
\be 
r\sin(\alpha\theta)=-\a\l_1,\quad r\sin(\alpha(\theta-\pi))=-\a\l_2 ,
\ee  
which intersect in the bulk at two positions $\theta=\theta_*, \theta_*+\pi$ and we will choose the shaded region in Figure \ref{bcc} as the bulk dual.
The intersection angle is not important to us, the point is that there always exists $T_{i=1,2}$ for such an intersection to happen. The on-shell action of this solution is given by
\be
I=-\frac{c \alpha^2}{24}\beta,
\ee 
which implies that the conformal dimension of the BCC operator is
\be 
h_{bcc}=\frac{c}{24}(1-\alpha^2),\quad \Delta_{bcc}=2h_{bcc}=\frac{c}{12}(1-\alpha^2).
\ee 
Note that for our choice of the bulk dual, the conical singularity at the center of AdS space is not included, so the BCC operator is solely dual to the defect which stays in the bulk. Assuming $\beta\rightarrow \infty$, then one can think that the BCC operators $\phi_i$ are inserted at $\tau=-\infty$ and $\tau=\infty$ such that the partition function is given by
\be 
Z=e^{-I}=\langle\phi|e^{-\beta(L_0-\frac{c}{24})}|\phi\rangle=e^{-\beta(h_{bcc}-\frac{c}{24})}.
\ee 
In this paper, we are interested in another type of gravity solutions where the defect can intersect with the AdS boundary. We expect that such solutions correspond to the correlation functions of BCC operators. First, let us recall some known facts about the correlation function of BCC operators.  Two-point function of BCC operators on a disk of radius $\beta$ (see figure \ref{EOW1}) is given by \cite{Almheiri:2021jwq} \footnote{For more discussions about the correlation functions of BCC operators, see \cite{affleck1994fermi}.}
\be \label{bcc_twopoint}
\langle \mathcal{O}_{J_1J_2}(0)\mathcal{O}_{J_2J_1}(\tau)\rangle_{\text{disk}}=\(\frac{\pi \epsilon}{\beta \sin(\pi\tau/\beta)}\)^{2\Delta[J_2,J_1]}
\ee 
which in the limit $\tau/\beta\rightarrow 0$ is approximately
\be \label{approxtwo}
\langle \mathcal{O}_{J_1J_2}(0)\mathcal{O}_{J_2J_1}(\tau)\rangle_{\text{disk}}\sim \(\frac{\epsilon}{\tau}\)^{2\Delta[J_2,J_1]},
\ee 
where $\Delta[J_2,J_1]$ is the conformal dimension of the operator $\mathcal{O}_{J_1J_2}$.
		
\begin{figure}[ht]
\centering
\includegraphics[width=0.45\textwidth]{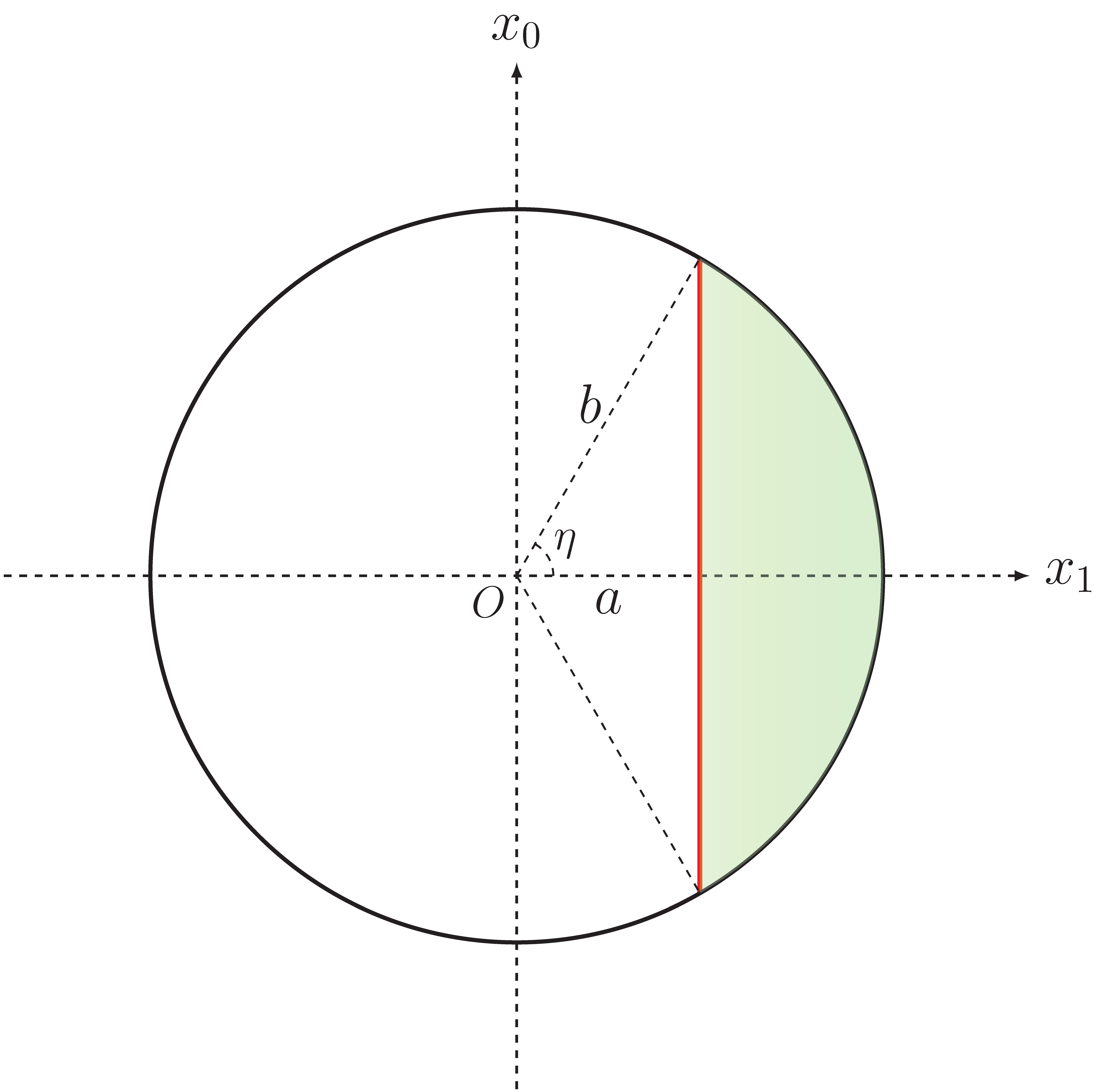}
\caption{The bulk region in \ref{EOW1} is dual to a BCFT defined on the green-shaded region..}
\label{EOW2}
\end{figure}

The simplest setup is the following. We consider the \Poincare AdS$_3$ bulk solution with metric
\bea\label{Poincare}
ds^2=\frac{dz^2+dx_0^2+dx_1^2}{z^2}
\eea
and two  (tensionless) EOW branes
\bea 
x_0^2+x_1^2+z^2= b^2,\quad x_1=a\equiv b\cos(\eta),\quad 0<\eta<\frac{\pi}{2}.
\eea 
The setup is illustrated in Figure \ref{EOW1} and \ref{EOW2}. 
At the AdS boundary, these two EOW branes intersect at $x_0=\pm \tau_0,\tau_0=b\sin(\eta)$. When $\eta$ is small: $\eta<<1$, the leading contribution to the on-shell action can be approximated by the Weyl anomaly \cite{henningson1998holographic}
\be 
I\approx \frac{2\eta}{8\pi G_N}\log\frac{\epsilon}{b}-\(-\frac{2\tan(\eta)}{8\pi G_N}\log\frac{b}{\tau_0}\)\approx \frac{\eta}{4\pi G_N}\log\frac{\epsilon}{\tau_0},
\ee  
leading to
\be 
Z=e^{-I}\approx \(\frac{\epsilon}{\tau_0}\)^{-2\Delta},\quad \Delta=\frac{c}{12}\frac{\h}{\pi}.
\ee 
This result is equal to the inverse of the two-point function \eqref{approxtwo}. In other words, the two-point function is dual to the \Poincare spacetime with this region excised. In the following sections, we will demonstrate that this phenomenon is quite common; specifically,  correlation functions can be computed from the on-shell action of an excised geometry. For the two-point function of two bulk primaries, this result can be elucidated using the wedge construction of the conical geometry as previously discussed.

\section{Cases study}
\label{sec:cases}
\renewcommand{\theequation}{3.\arabic{equation}}
\setcounter{equation}{0}
In this section we construct the excised geometry on top of various AdS$_3$ solutions including the global AdS, \Poincare AdS and the BTZ black holes. Our construction strategy is as follows: we begin with the global AdS geometry  described by the metric \eqref{global_metric} and excise a wedge defined by\footnote{In the construction of the conical geometry without boundaries in the bulk, we also need to glue the two edges of the wedge.}
\be
 -\h<\theta<\h, \quad \h=\pi\(1-\a\)=4\pi G_Nm.
 \ee

As previously explained, this excised geometry is dual to a conical geometry, with the tip of the wedge corresponding to a massive particle situated at the center of the global AdS. Next, we perform a boost transformation \eqref{boost global}. In this boosted frame, the trajectory of the particle (or the tip of the wedge) intersects the AdS boundaries at specific points, and the wedge now resembles a spindle. It is important to note that the boost transformation is an isometry of the global AdS spacetime, thus, we have constructed a global AdS spacetime with a spindle-like region excised, which is expected to be dual to the two-point function of two scalar primaries in a CFT defined in a finite interval. This resulting excised solution has already been considered in \cite{Caputa:2022zsr} as a holographic description of the Virasoro coherent state. To construct other excised geometries, we leverage the fact that there are no local degrees of freedom, allowing us to use the known coordinate transformation to transform one excised geometry into another. This strategy is quite standard, for example, it has been employed in the construction of shockwave geometries \cite{aichelburg1971gravitational,dray1985effect,dray1985gravitational,hotta1993shock,Sfetsos:1994xa}  caused by the backreaction of the massless particle.  After constructing the excised geometries, we will compute the on-shell action either analytically or numerically to demonstrate that they indeed yield the correct two-point functions. It is important to emphasize that we are performing an honest bulk calculation, meaning that we compute the on-shell actions of the excised geometries directly, rather than transforming them into other simple configurations as done in \cite{Caputa:2022zsr}, which only matches our results at leading order.

\subsection{\Poincare AdS$_3$}
\begin{figure}[htbp]
	\centering
	\includegraphics[width=0.55\textwidth]{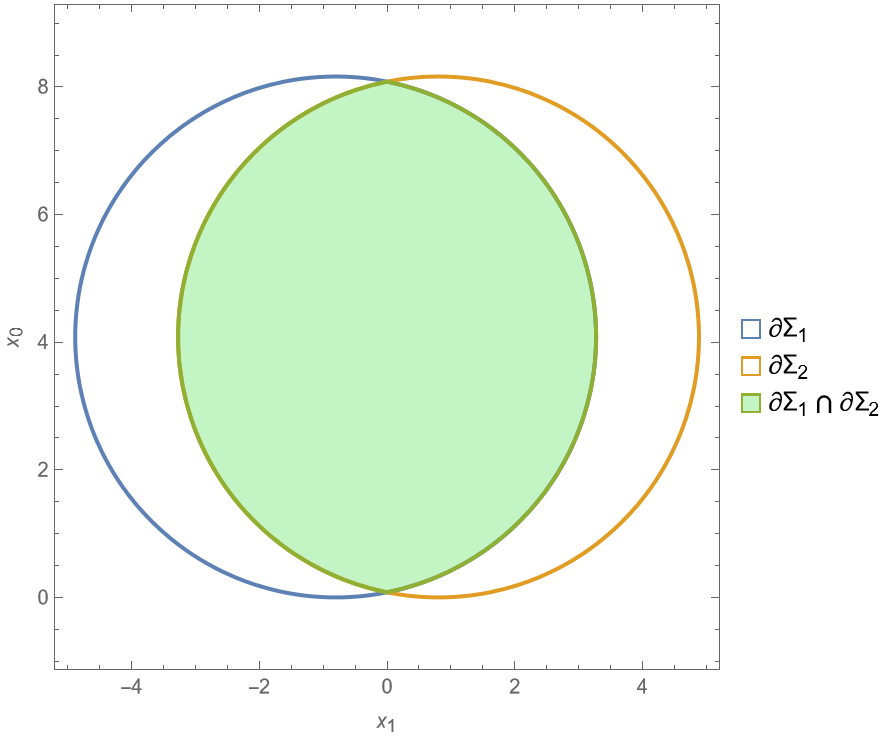}
	\caption{Boundary region dual to the bulk wedge in \Poincare AdS$_3$.}
	\label{fig: Poincare-1}
\end{figure}

We begin by examining the simplest case: the excised \Poincare AdS$_3$, where all computations can be performed exactly. Given the metrics \eqref{global_metric} and \eqref{Poincare} of the global AdS$_3$ and the \Poincare AdS$_3$, we make the following identifications:
\bea\label{boost poincare}
\sqrt{r^2+1}\cosh\tau &=& \cosh\b \frac{z^2+x_1^2+x_0^2+1}{2z}-\sinh\b \frac{x_0}{z},\nonumber\\
r\cos\q &=& -\sinh\b \frac{z^2+x_1^2+x_0^2+1}{2z}+\cosh\b \frac{x_0}{z},\nonumber\\
r\sin\q  &=& \frac{x_1}{z}, \nonumber\\
\sqrt{r^2+1}\sinh\tau &=&\frac{z^2+x_1^2+x_0^2-1}{2z}.
\eea
It should be pointed out that when the boost parameter $\beta=0$, the transformations \eqref{boost poincare} reduce to the standard coordinate transformations between global AdS$_3$ and \Poincare AdS$_3$. 
These identifications imply that  the trajectory of a static particle in the global AdS$_3$ maps to a circle described by
\bea\label{worldline}
\(x_0-\coth\b\)^2+z^2=\frac{1}{\sinh^2\b}, \quad x_1=0\, ,
\eea
which intersects the AdS asymptotic boundary at the following two points
\be\label{insertion poincare}
x_{0,a}, \; x_{0,b}=\coth\frac{\b}{2},\; \tanh\frac{\b}{2}.
\ee
Meanwhile, the edges of the wedge, $\q=\pm\eta$, are mapped into circles on a sphere defined by
\be\label{surfaces Poincare}
\S_{1,2}: \(x_0-\coth\b\)^2+\(x_1\pm\frac{1}{\sinh\b \tan\h}\)^2+z^2=\frac{1}{\sin^2\h\sinh^2\b},
\ee
which represent the profiles of the tensionless EOW branes in \Poincare AdS$_3$. Notably, the intersection of these surfaces is precisely the worldline given in \eqref{worldline}. According to AdS/BCFT correspondence, the wedge $\Sigma_1\cap\Sigma_2$ is dual to a BCFT defined on $\pa\S_{1} \cap \pa\S_{2}$, as illustrated in Figure \ref{fig: Poincare-1}.

The on-shell action \eqref{BCFTaction} of this excised geometry has been computed in \cite{Caputa:2022zsr} and is also reproduced in Appendix \ref{append:A}. Here, we provide some comments on this computation. The on-shell action in the total \Poincare AdS$_3$ is zero, as it corresponds to the vacuum state of a CFT defined in the complex plane. Therefore, the on-shell action of the excised \Poincare geometry is equal to $-I_{\text{gravity},W}$. As explained in Section \ref{sec:wedge_proposal}, the matter term cancels with the Hayward term associated with $\partial\Sigma_1\cap \partial\Sigma_2$. The other two Hayward terms related to $\partial\Sigma_i\cap \partial \rm AdS$ contribute only finite amounts. Consequently, the leading order contribution of \eqref{BCFTaction} arises from the Weyl anomaly of the dual BCFT. The result is given by:
\bea
I_{grav}=-I_{gravity,W}=-\frac{1}{8\pi G_N}\[4\h\log\frac{\e}{b}-\im\(\text{Li}_{2}(e^{2\im\h})-\text{Li}_{2}(e^{-2\im\h})\)-4\h\]
\eea
Recall the relation between the conformal dimension and the local mass of the particle is given by
\be
\Delta=m\(1-2G_N m\)\approx m, \quad \text{when}\quad G_Nm\ll 1,
\ee
thus in the probe limit the partition function of the excised geometry exactly reproduces the two-point function
\be
e^{-I_{grav}}\approx \frac{\e^{2h+2\bar{h}}}{(u_{12})^{2h}\;(v_{12})^{2\bar{h}}}
=\(\frac{\e}{x_{0,ab}}\)^{2\D}.
\ee
where $h=\bar{h}=\frac{\D}{2}$ for scalar primary operators and the light-cone coordinate
\be
u=x_{1}+\im x_0,\quad v=x_{1}-\im x_0
\ee
is adopted.

\subsection{Global AdS$_3$}
\begin{figure}[htbp]
	\centering
	\includegraphics[width=0.55\textwidth]{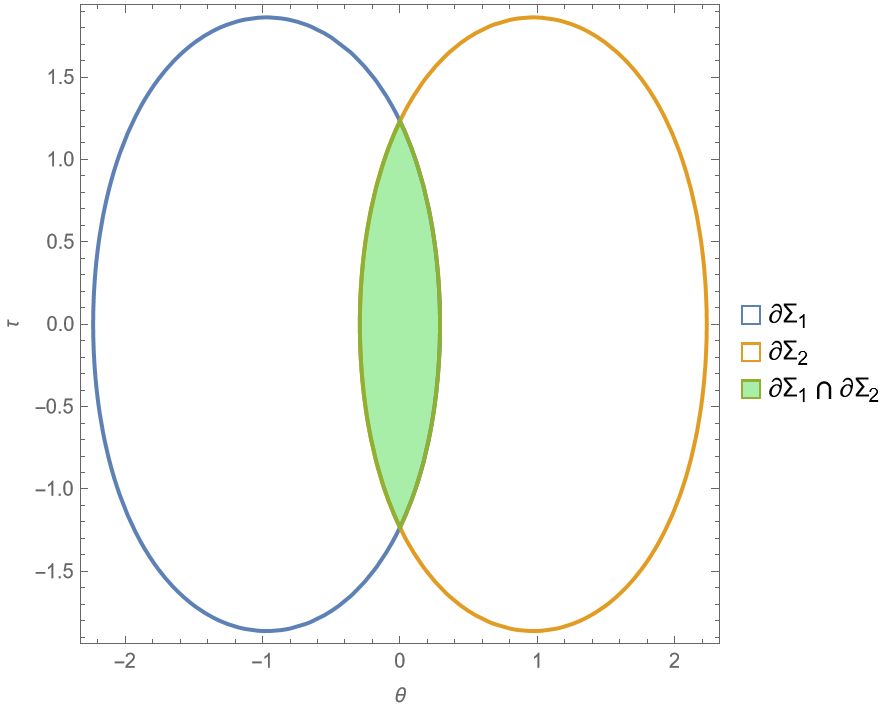}
	\caption{Boundary region dual to the bulk wedge in global AdS$_3$.}
	\label{fig:wedge gAdS3}
\end{figure}

In global AdS$_3$, the construction of the excised geometry is straightforward. We need only to perform the following boost:
\bea\label{boost global}
\sqrt{r_{s}^2+1}\cosh\t_{s}&=&\sqrt{r^2+1}\cosh\(\tau-\tau_0\)\cosh\b-r \cos\(\q-\q_0\)\sinh\b\nonumber\\
r_{s}\cos\q_{s}&=&-\sqrt{r^2+1}\cosh\(\tau-\tau_0\)\sinh\b+r\cos\(\q-\q_0\)\cosh\b\nonumber\\
r_{s}\sin\q_{s}&=&r\sin\(\q-\q_0\)\nonumber\\
\sqrt{r_{s}^2+1}\sinh\t_{s}&=&\sqrt{r^2+1}\sinh\(\tau-\tau_0\)
\eea
where we have added a subscript $s$ to denote the static frame for  distinction. We can solve for the profile of the moving particle by setting $r=0$ in the above transformations. The result is
\be
\frac{r}{\sqrt{r^2+1}}=\tanh\b\cosh\(\tau-\tau_0\),\quad \theta=\theta_0+k\pi, k\in\mathbb{Z}.
\ee
which corresponds precisely to the intersection of two tensionless EOW branes with the profiles
\bal\label{branes gAdS3}
\S_{1,2}: \pm\tan\h=\frac{r\sin\(\q-\q_{0}\)}{\sqrt{r^2+1}\cosh\(\tau-\tau_{0}\)\sinh\b-r\cos\(\q-\q_{0}\)\cosh\b}.
\eal
Without loss of generality, we set $k=\theta_0=\tau_0=0$ in the following calculations. The two insertion points at the boundary are given by
\bal\label{global-insertion}
\tau_{1}, \; \tau_{2}=\pm \arccosh\(\coth\b\)=\pm\tau_{b},\quad \coth\b\equiv\cosh\tau_{b}.
\eal
The relationship between the coordinate times defined along the worldline of the particle in the static and boost frame is
\bal
\tanh\t_{s}=\cosh\b\tanh\tau.
\eal 
As shown in Figure \ref{fig:wedge gAdS3}, the excised wedge also takes the shape of a spindle, which is dual to a BCFT defined in the region $\partial\Sigma_1\cap \partial\Sigma_2$ with 
\bea
\pa\S_{1,2}: \pm\tan\h=\frac{\sin\q}{\cosh\tau\sinh\b-\cos\q\cosh\b}.
\eea

Similar to the previous case, the on-shell action of the entire global AdS$_3$ geometry and the Hayward terms only give finite contributions. The leading order contribution to the on-shell action arises from that in the wedge:
\bea
I_{gravity, W}&=&-\frac{1}{16\pi G_N}\int_{\cM}\sqrt{g}\(R+2\)-\frac{1}{8\pi G_N}\int_{\pcM}\sqrt{h}\(K-1\)\nonumber\\
&=&-\frac{1}{8\pi G_N}\int_{\pcM} d^{2}x\(r^{2}(\q,\t)+\frac{1}{2}\),
\eea
where $r(\theta,\tau)$ is a function of $\theta$ and $\tau$ determined from \eqref{branes gAdS3}. To regularize the on-shell action, we choose a standard cutoff $r=\frac{1}{\epsilon}$. Then the integration domain $\partial\Sigma_1\cap \partial \Sigma_2$ is explicitly given by:
\be
-\q(\tau)<\q<\q(\tau), \quad -\tau_{c}<\tau<\tau_{c},\quad \tau_{c}=\arccosh\frac{\coth\b}{\sqrt{1+\e^2}}.
\ee
Here, $\q$ is a function of $\tau$, determined from 
\eqref{branes gAdS3}, yielding
\bea
\sin\(\q(\tau)-\Q_0\)=-\m\sqrt{1+\e^2}
\eea
where we have introduced the following variables: 
\be
\cosh\b\tan\h=\tan\Q_0,\quad\m=\tanh\b\sin\Q_0\cosh\t,
\ee
for convenience. Since the wedge region is symmetric about $\q=0$ and $\tau=0$, we will only consider the case $0<\h<\frac{\pi}{2}$, which implies $0<\Q_0<\frac{\pi}{2}$ as $\b$ is assumed to be positive. 

\begin{figure}[htbp]
	\centering
	\includegraphics[width=0.75\textwidth]{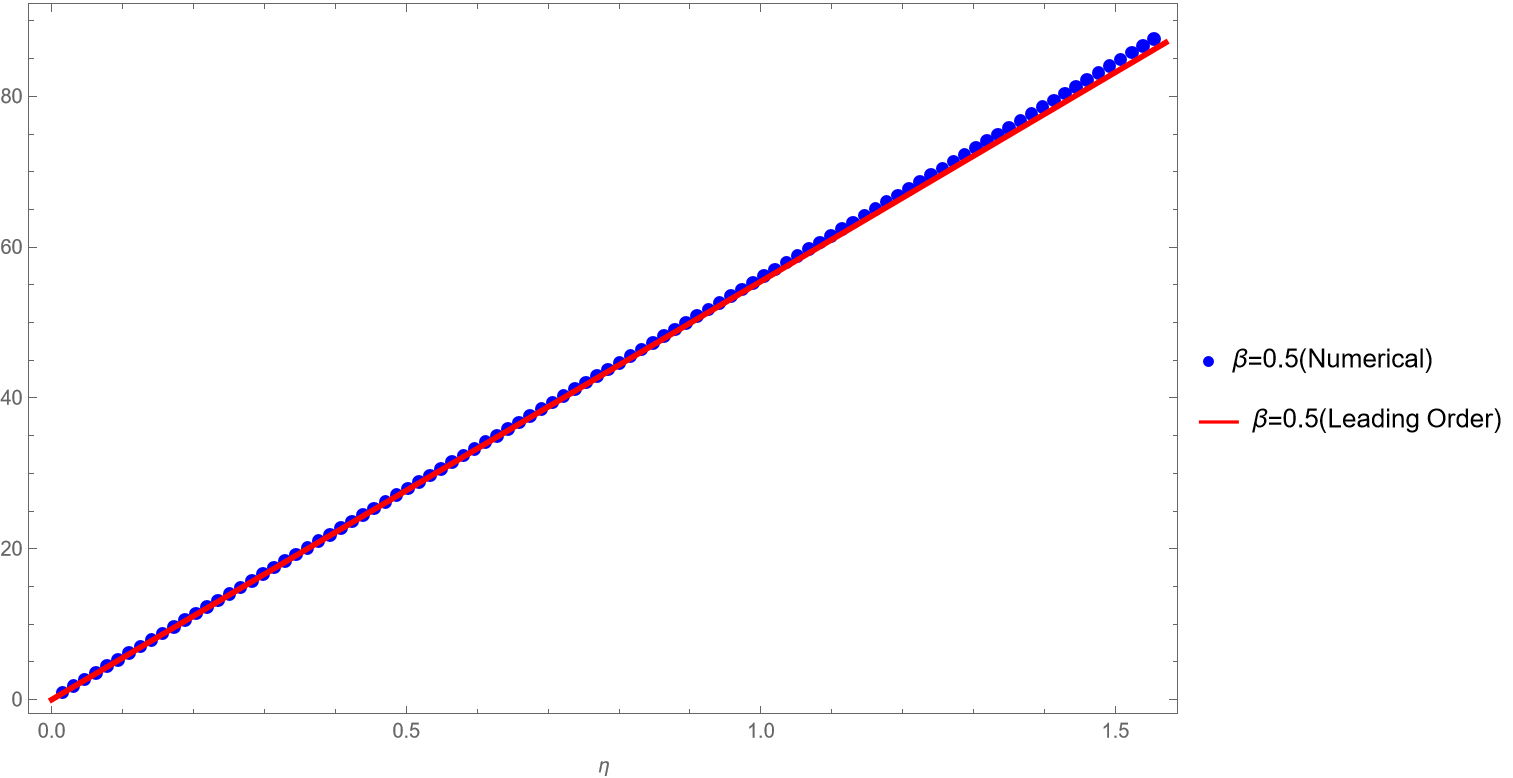}
	\caption{Numerical result of the integral in global AdS$_3$ when $\b=0.5,\e=10^{-6}$.}
	\label{fig:global_num}
\end{figure}

Substituting the explicit expression $r(\theta,\tau)$ into this integral gives			
\bea
I_{gravity, W}&=&-\frac{1}{8\pi G_N}\int_{-\t_{c}}^{\t_{c}}d\t\int_{0}^{\q(\t)}d\q\,\frac{\(\cos\q\cosh\b\tan\h-\sin\q\)^2+\(\tan\h\sinh\b\cosh\t\)^2}{\(\cos\q\cosh\b\tan\h-\sin\q\)^2-\(\tan\h\sinh\b\cosh\t\)^2}\nonumber\\
&=&-\frac{1}{8\pi G_N}\int_{-\t_{c}}^{\t_{c}}d\t\int_{0}^{\q(\t)}d\q\[1+\frac{2\m^2}{\sin^2\(\q-\Q_0\)-\m^2}\]\nonumber\\
&=&-\frac{1}{8\pi G_N}\int_{-\tau_c}^{\t_{c}}d\t\[\q\(\tau\)+\frac{\m\(\log\frac{4\(1-\m^2\)}{\e^2}-\log\frac{\tan\Q_0+\frac{\m}{\sqrt{1-\m^2}}}{\tan\Q_0-\frac{\m}{\sqrt{1-\m^2}}}\)}{\sqrt{1-\m^2}}\].\label{global_wedge_action}
\eea
Note that the first term that appears in the integral is proportional to the area of the region $\pa\S_{1}\cap \pa\S_{2}$, corresponding to the vacuum energy of the bulk wedge region. The second term plays a role similar to that in the case of the \Poincare AdS solution, contributing to the Weyl anomaly of the BCFT. The integral is too complicated to evaluate analytically. A numerical result is shown in Figure \ref{fig:global_num}. We observe a linear dependence with respect to the angle $\eta$ when $\eta$ is small.  A similar linear dependence is also observed in the case of \Poincare AdS solution. Motivated by these facts, to have some analytic understanding of the integral,  we compute this integral perturbatively, order by order with respect to the deficit angle $\h$. Expanding the integrand to the leading order of $\eta$ gives
\bea
\h\(\cosh\b-\sqrt{1+\e^2}\sinh\b\cosh\tau\)-\h\sinh\b\(\log\frac{\e^2}{4}+\log\frac{\coth\b+\cosh\tau}{\coth\b-\cosh\tau}\)\cosh\tau.\quad
\eea 
and  evaluating the resulting integral over $\tau$ gives the leading result of \eqref{global_wedge_action}:
\bea
-\frac{1}{8\pi G_N}\[-2\h\log\frac{\e^2}{4\sin\frac{w_{12}}{2}\sin\frac{\bar{w}_{12}}{2}}-2\h\(1+\cosh\b\log\(\coth\frac{\b}{2}\)\)\]
\eea
where the light-cone coordinates have been introduced
\be
w_{1}=\q_{1}+\im \tau_{1}=\im \tau_{b},\quad w_{2}=\q_{2}+\im \tau_{2}=-\im \tau_{b}.
\ee

As shown in the Figure \ref{fig:global_num}, the analytical result at leading order is consistent with the numerical result when $\b$ or $\h$ is small. Once we include the next order corrections, the deviation approaches zero as shown in Appendix \ref{append:C} for $\beta$ and $\eta$ in a very large domain. Therefore, we conclude that the partition function is given by:
\be
I_{grav}=-I_{gravity,W}\approx m\log\frac{\e^2}{4\sin\frac{w_{12}}{2}\sin\frac{\bar{w}_{12}}{2}}+m\(1+\cosh\b\log\(\coth\frac{\b}{2}\)\)
\ee
which implies
\bea
\frac{e^{-I_{grav}}}{e^{-I_{grav},vacuum}}\approx\frac{\e^{2(h+\bar{h})}}{\(2\sin\frac{w_{12}}{2}\)^{2h}\times\(2\sin\frac{\bar{w}_{12}}{2}\)^{2\bar{h}}}
\eea
with $h=\bar{h}=\frac{m}{2}$ for scalar primary operators in the probe limit. It precisely reproduces the two-point function in a CFT defined on a finite circle.
		

\subsection{BTZ}
Finally, let us construct the excised BTZ geometry. The BTZ geometry \cite{Banados:1992gq,Banados:1992wn} can be described by the metric
\be
ds^2=\frac{f(z)d\t^2}{z^2}+\frac{dz^2}{f(z)z^2}+\frac{dx^2}{z^2},\quad f(z)=1-\frac{z^2}{z_H^2},\quad \tau\sim \tau+2\pi z_H.
\ee 
Assuming rotational symmetry in its trajectory, the geodesic equation in the BTZ geometry can be solved as
\be\label{geodesicBTZ}
z(\t)=z_H\sqrt{1-\(1-\frac{z_0^2}{z_H^2}\)\(1+\tan^2\frac{\t-\t_0}{z_H}\)}
\ee
where $\t_0$ and $z_0$ are two integral constants. Below, we will obtain this by using our aforementioned strategy as the intersection of two tensionless EOW branes in the boosted frame. Recalling that the coordinates $(\tau,z,x)$ are related to the embedding coordinates via
\bea\label{btz}
&&X_0=\frac{z_H}{z}\cosh\frac{x}{z_H}, \quad  X_1=\sqrt{\frac{z_H^2}{z^2}-1}\cos\frac{\t}{z_H},\nn
&&X_2=\frac{z_H}{z}\sinh\frac{x}{z_H}, \quad  X_3=\sqrt{\frac{z_H^2}{z^2}-1}\sin\frac{\t}{z_H},
\eea
we find that the boosted frame should be defined by 
\bea\label{boost btz}
\sqrt{r^2+1}\cosh\tilde{\t}&=&\frac{z_H}{z}\cosh\frac{x}{z_H}\cosh\b- \sqrt{\frac{z_H^2}{z^2}-1}\cos\frac{\t}{z_H}\sinh\b,\\\nonumber
r\cos\q&=&-\frac{z_H}{z}\cosh\frac{x}{z_H}\sinh\b+ \sqrt{\frac{z_H^2}{z^2}-1}\cos\frac{\t}{z_H}\cosh\b,\\\nonumber
r\sin\q&=&\frac{z_H}{z}\sinh\frac{x}{z_H},\\\nonumber
\sqrt{r^2+1}\sinh\tilde{\t}&=&\sqrt{\frac{z_H^2}{z^2}-1}\sin\frac{\t}{z_H},
\eea
where $\tilde{\tau}$ has been introduced to denote the time in the global AdS$_3$ for dinstinction. It should be pointed out that a similar relation was used in \cite{Caputa:2014eta,Stikonas:2018ane} for studying the entanglement entropy of the (rotating) BTZ black hole perturbed by a massive backreacted free-falling particle. The relations \eqref{boost btz} imply that the surfaces $\q=\pm\h$ are mapped into two EOW brane profiles
\bea\label{btzwedge surfaces}
\S_{1,2}: \pm\tan\h=\frac{\sinh\frac{x}{z_H}}{-\sinh\b\cosh\frac{x}{z_H}+\sqrt{1-\frac{z^2}{z_H^2}}\cos\frac{\t}{z_H}\cosh\b}.
\eea
The intersection is given by
\be
z=z_H\sqrt{1-\tanh^2\b\(1+\tan^2\frac{\t}{z_H}\)},\quad x=0,
\ee
which coincides with the geodesic \eqref{geodesicBTZ}, up to the following identification 
\be
z_0=\frac{z_H}{\cosh\b},\quad \t_0=0.
\ee
Note that $z_0$ is also the deepest position of this geodesic in the bulk, which is outside the horizon. The transformations \eqref{boost btz} also imply that the two endpoints of this geodesic are at 
\bea
\t_{1,2}=\pm z_H\tau_b,\quad x_{1,2}=0,\quad \tanh\b\equiv\cos\t_b,
\eea
and the relation between the coordinate time along the trajectory in the boosted frame and the static frame is
\be
\tanh\tilde{\t}=\sinh\b\tan\frac{\t}{z_H}.
\ee
\begin{figure}[htbp]
	\centering
	\subfigure[$\mathcal{G} < 1$]
	{
		\begin{minipage}[b]{0.47\linewidth}
			\includegraphics[width=1\linewidth]{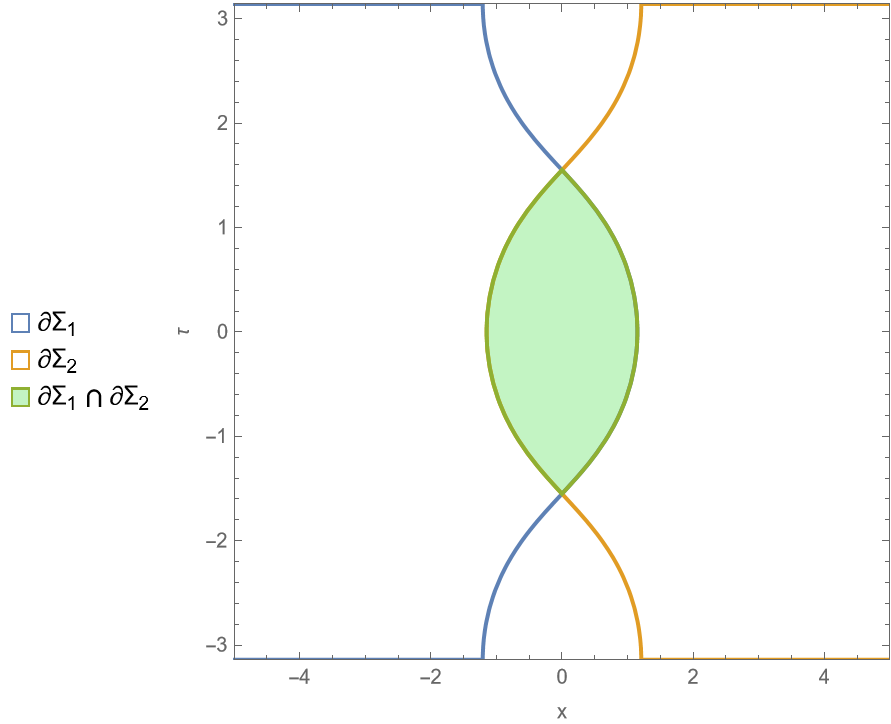}
		\end{minipage}
		\label{fig:btz-1}
	}
	\subfigure[$\mathcal{G} > 1$]
	{
		\begin{minipage}[b]{0.48\linewidth}
			\includegraphics[width=1\textwidth]{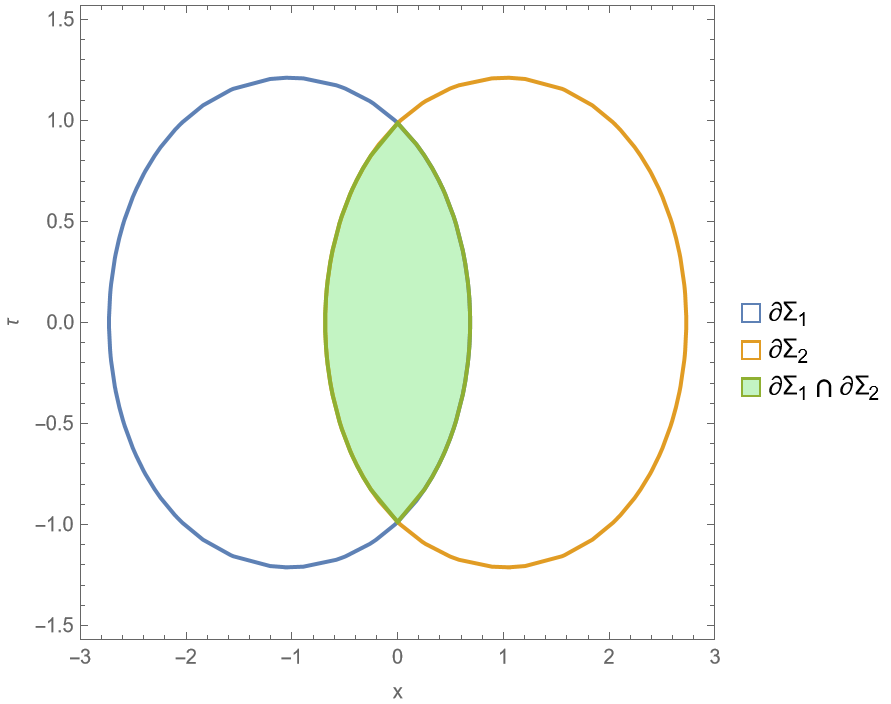}
		\end{minipage}
		\label{fig:btz-2}
	}
	\caption{Boundary region dual to the bulk wedge in BTZ black hole}
	\label{fig:btzwedge}
\end{figure}
As shown in Figure \ref{fig:btzwedge}, the wedge in the boosted frame still has the shape of a spindle, but the EOW branes have two possible configurations depending on the value of the combination $\mathcal{G}=\tan\h\sinh\b$. When $0< \mathcal{G} \leq 1$, the surfaces $\S_{1,2}$  terminate at the event horizon $z=z_{H}$; whereas, when $\mathcal{G} > 1$, the surfaces $\S_{1,2}$ will turn around before reaching the horizon,  resembling the behavior in global AdS$_3$. In either case, the dual BCFT is defined in  $\partial \Sigma_1\cap \partial\Sigma_2$ with \be
\pa\S_{1,2}: \pm\tan\h=\frac{\sinh\frac{x}{z_H}}{-\sinh\b\cosh\frac{x}{z_H}+\cos\frac{\t}{z_H}\cosh\b}.
\ee

\subsubsection*{on-shell action: $\mathcal{G}<1$}		
First, let us consider the case of $\mathcal{G}<1$. As explained before, the leading contribution to the on-shell action comes from that in the wedge:
\bea
I_{gravity,W}&=&-\frac{1}{16\pi G_N}\int_{\cM}\sqrt{g}\(R+2\)-\frac{1}{8\pi G_N}\int_{\pcM}\sqrt{h}\(K-1\)   \nn
&=& \frac{1}{4\pi G_N}\int_{\pcM}d^{2}x\(\frac{1}{2\e^2}-\frac{1}{2z^2\(\tau,x\)}\)-\frac{1}{8\pi G_N}\int_{\pcM}d^{2}x\(\frac{1}{\e^2}-\frac{1}{2z_H^2}\)\nn
&=&-\frac{1}{8\pi G_N}\int_{\pcM} d^{2}x\(\frac{1}{z^2\(\tau,x\)}-\frac{1}{2z_H^2}\). \label{BTZ_integral}
\eea
where $z$ is a function of $x$ and $\tau$ solved from \eqref{btzwedge surfaces}. Then the integration domain $\partial\Sigma_1\cap \partial \Sigma_2$ is explicitly given by:
\bea \label{BTZ_domain}
&&-x(\tau)<x<x(\tau),\quad -\tau_{c}<\tau<\tau_{c},\\
&&x(\tau)+x_0=z_H\sinh^{-1}\(\m\sqrt{1-\frac{\e^2}{z_H^2}}\),\quad \tau_{c}=z_H\arccos\frac{\tanh\b}{\sqrt{1-\frac{\e^2}{z_H^2}}},
\eea 
where we have chosen a  cutoff $z=\e$ for regularization and introduced the following convenient variables: 
\bea
\tanh\frac{x_0}{z_H}=\mathcal{G}\equiv\tan\h\sinh\b, \quad\mu=\sinh\frac{x_0}{z_H}\coth\b\cos\frac{\tau}{z_H}.
\eea
\begin{figure}[htbp]
	\centering
	\includegraphics[width=0.75\textwidth]{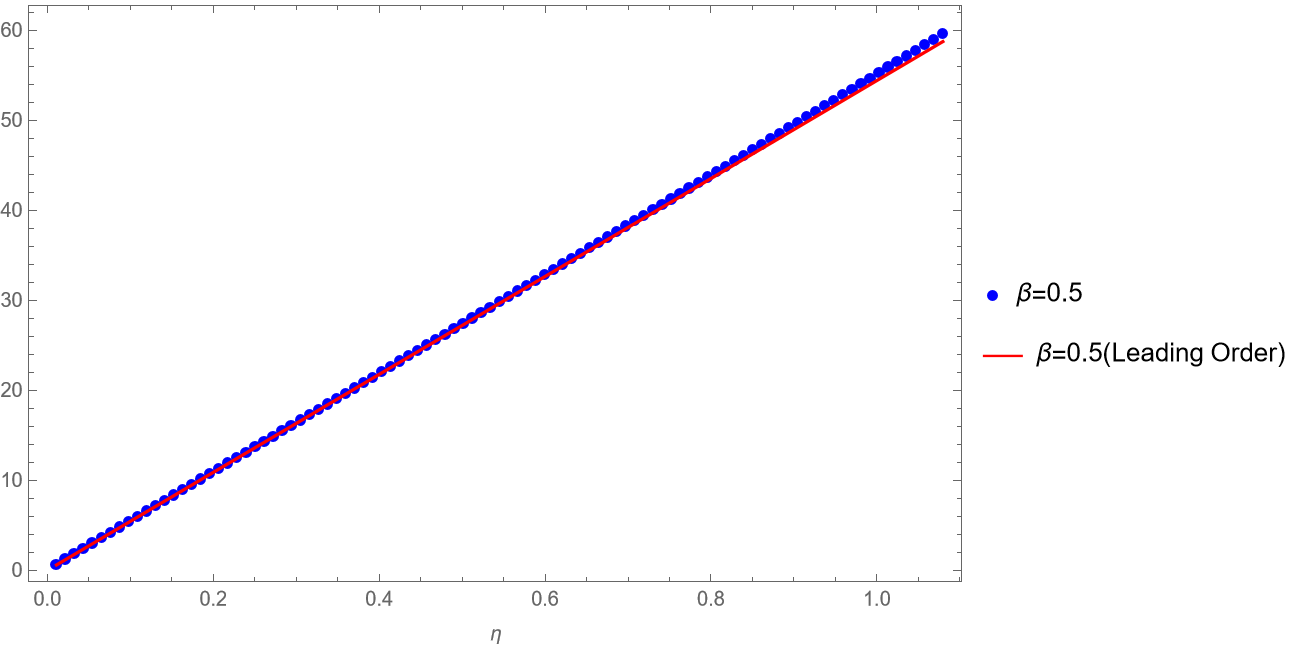}
	\caption{Numerical result and analytical result at leading order perturbation when $\mathcal{G}<1$.}
	\label{numerical:btzi}
\end{figure}
Substituting the explicit expression of $z(\tau,x)$ into the above integral, we obtain
\bea
I_{gravity,W}&=&-\frac{1}{8\pi G_N z_H^2}\int_{-\tau_{c}}^{\tau_{c}}d\tau\int_{0}^{x(\tau)}\(-1+\frac{2\m^2}{\m^2-\sinh^2\frac{x+x_0}{z_H}}\)dx\nn
&=&-\frac{1}{8\pi G_N z_H}\int_{-\tau_{c}}^{\tau_{c}}d\tau\[-\frac{x\(\tau\)}{z_H}+\frac{\m\(\log\frac{4z_H^2\(1+\m^2\)}{\e^2}-\log\frac{\frac{\m}{\sqrt{\m^2+1}}+\tanh\frac{x_0}{z_H}}{\frac{\m}{\sqrt{\m^2+1}}-\tanh\frac{x_0}{z_H}}\)}{\sqrt{\m^2+1}}\].\quad\quad\quad   
\eea
To tackle this integral, we employ our previous perturbative strategy. When $\h$ is small and the parameter $\b$ satisfies $\b<\sinh^{-1}\(\cot\h\)$, we can expand the integrand in terms of $\h$, at the first order yielding
\bea
\h\(\sinh\b-\sqrt{1-\frac{\e^2}{z_H^2}}\cos\frac{\tau}{z_H}\cosh\b\)+\h\cosh\b\(\log\frac{4z_H^2}{\e^2}-\log\frac{\cos\frac{\tau}{z_H}+\tanh\b}{\cos\frac{\tau}{z_H}-\tanh\b}\)\cos\frac{\tau}{z_H}.\nonumber
\eea 
Now, this integral can be analytically evaluated, and the result is 
\bea
I_{gravity,W} \approx m\(\log\frac{\e^2}{4z_H^2\sinh\frac{w_{12}}{2z_H}\sinh\frac{\bar{w}_{12}}{2z_H}}+1+\cos^{-1}\(\tanh\b\)\sinh\b\)=-I_{grav},
\eea
where we have used identity $\sinh\frac{w_{12}}{2z_H}\sinh\frac{\bar{w}_{12}}{2z_H}=\sech^2\b$, $\eta=4\pi G_N m$, and introduced light-cone coordinates
\bea
w_{1}=x_{1}+\im\tau_{1}=\im z_H\tau_{b},\quad w_{2}=x_{2}+\im\tau_{2}=-\im z_H\tau_{b}. 
\eea As illustrated in Figure \ref{numerical:btzi}, when $\h<1$, the numerical result is almost linear to $\h$ and nearly coincides with the analytical result at leading order perturbation. Once  the deficit angle $\h$ is increased, the contribution of high-order perturbation should be included so that the deviation will be smoothed, as illustrated in the appendix \ref{append:C}. 

Since the second term in the above result is nearly equal to the boundary area of the BCFT region, if we only consider the leading order contribution, the partition function of the excised geometry exactly reproduces the two-point function in BTZ black hole
\bea
\frac{e^{-I_{grav}}}{e^{-I_{grav,vacuum}}}\approx \frac{\e^{2h+2\bar{h}}}{\(2z_H\sinh\frac{w_{12}}{2z_H}\)^{2h}\(2z_H\sinh\frac{\bar{w}_{12}}{2z_H}\)^{2\bar{h}}}
\eea
with $h=\bar{h}=\frac{\D}{2}$ for scalar primary operators in the probe limit. It precisely reproduces the two-point function in a CFT in a thermal state.

\subsubsection*{on-shell action: $\mathcal{G}=1$}		
\begin{figure}[htbp]
	\centering
	\includegraphics[width=0.75\textwidth]{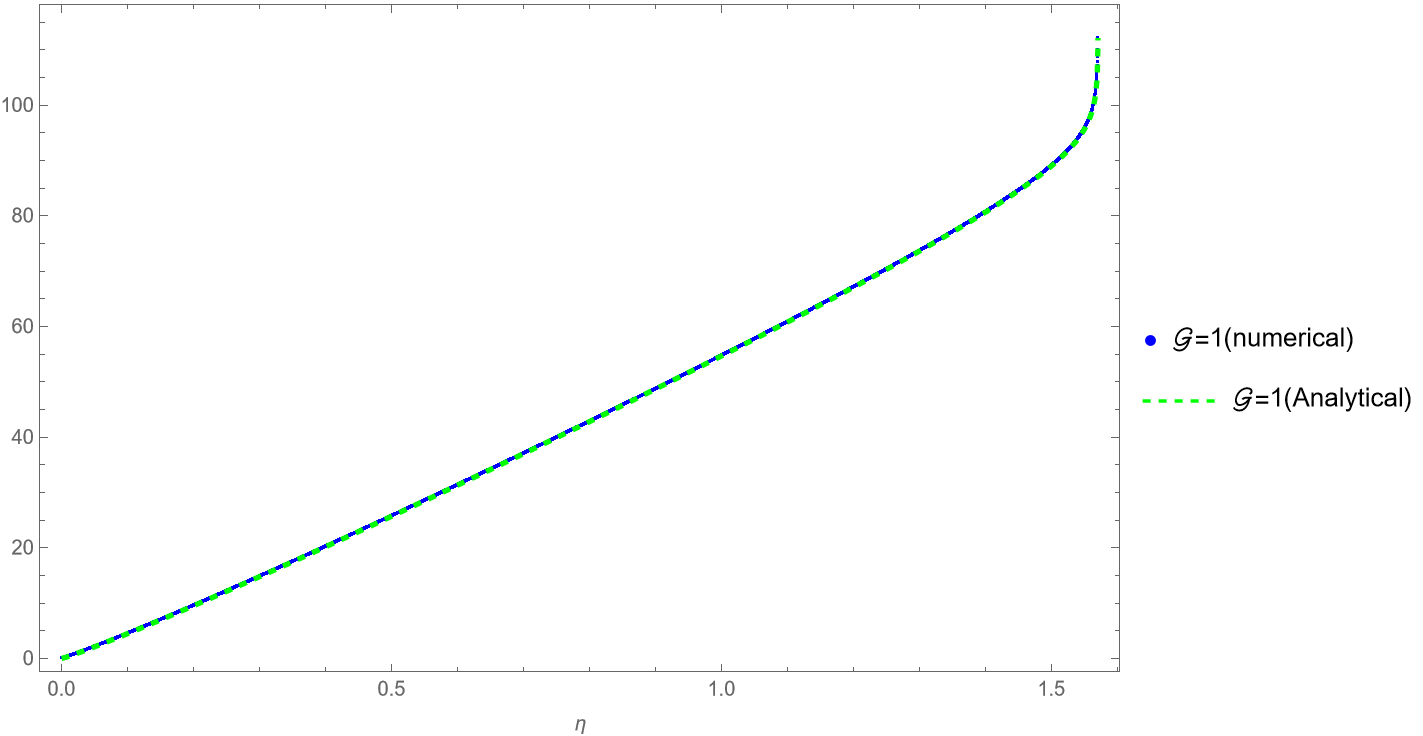}
	\caption{Comparison between the numerical result and analytical result when $\mathcal{G}=1$.}
	\label{numerical:btzii}
\end{figure}

In this case, the integral \eqref{BTZ_integral} is dramatically simplified to
\bea
I_{gravity,W}
&=&-\frac{1}{8\pi G_N z_H}\int_{-\tau_{c}}^{\tau_{c}}d\tau\[\frac{x\(\tau\)}{z_H}+\log\(\m^2-1\)-\log\frac{\m^2\e^2}{z_H^2}\]\nonumber\\
&=&-\frac{1}{8\pi G_N z_H}\int_{-\tau_{c}}^{\tau_{c}}d\tau\[-\log\frac{\e^2}{z_H^2}+\log\frac{\m^2-1}{\m}\]\nonumber\\
&=&-\frac{1}{8\pi G_N }\[-2\h\log\frac{\e^2}{z_H^2}+\mathcal{F}\(\frac{\t_{c}}{z_H}\)\],
\eea
where we have used the following expressions:
\bea  
&&x(\tau)=z_H\log\(\m\sqrt{1-\frac{\e^2}{z_H^2}} \,\),\quad  \m=\coth\b\cos\frac{\tau}{z_H},\nn
&&\frac{\tau_{c}}{z_H}=\arccos\(\frac{\cos\h}{\sqrt{1-\frac{\e^2}{z_H^2}}}\).
\eea   
Here, the complicated function $\mathcal{F}$ is given by
\footnote{Here, we have used the following result: $g(x)\equiv\int\log(\cos x)dx=-\frac{\text{i}x^2}{2}-x\log 2-\frac{\text{Li}_{2}(-e^{2\text{i}x})}{2\text{i}}$}
\bea
\mathcal{F}\(\frac{\t_{c}}{z_H}\)&=&\int_{-\frac{\t_{c}}{z_H}}^{\frac{\t_{c}}{z_H}}\log(\cos x) d x+\int_{-\frac{\t_{c}}{z_H}}^{\frac{\t_{c}}{z_H}}\log\(1-\cos^2\h\sec^2 x\)dx -\frac{2\t_{c}}{z_H}\log(\tanh\b)\nn
&=&-2\h\log(2\cos\h)+\frac{\im \(2\text{Li}_{2}(e^{2\im\h})-2\text{Li}_{2}(e^{-2\im\h})+\text{Li}_{2}(e^{4\im\h})-\text{Li}_{2}(e^{-4\im\h})\)}{4}.\quad\qquad 
\eea
As shown in Figure \ref{numerical:btzii}, this analytical result matches the numerical result very well.
When $\h$ is small, we find that in the leading order the on-shell action is 
\bea
I_{grav}&=&-I_{gravity,W}\approx-m\(\log\frac{\e^2}{4z_H^2\h^2}+2\)\approx-m\(\log\frac{\e^2}{4z_H^2\sin^2(\h)}+2\)\nn
&\approx& -m\(\log\frac{\e^2}{4z_H^2\sinh\frac{w_{12}}{2z_H}\sinh\frac{\bar{w}_{12}}{2z_H}}+2\).
\eea
This shows that, up to a redefinition of the cutoff surface, the partition function $e^{-I_{grav}}\sim e^{I_{gravity,W}}$ exactly reproduces the two-point function in BTZ geometry.

\subsubsection*{on-shell action: $\mathcal{G}>1$}		
In the end, let us consider the case of $\mathcal{G}>1$. The integral \eqref{BTZ_integral} is explicitly given by
\bea
I_{gravity,W}&=&-\frac{1}{8\pi G_N z_H^2}\int_{-\tau_{c}}^{\tau_{c}}d\tau\int_{0}^{x(\tau)}\(-1+\frac{2\m^2}{\m^2-\cosh^2\frac{x+x_0}{z_H}}\)dx\nn
&=&-\frac{1}{8\pi G_N z_H}\int_{-\tau_{c}}^{\tau_{c}}d\tau\[\frac{-x\(\tau\)}{z_H}+\frac{\m\(\log\frac{4z_H^2\(\m^2-1\)}{\e^2}-\log\frac{\frac{\sqrt{\m^2-1}}{\m}+\tanh\frac{x_0}{z_H}}{\frac{\sqrt{\m^2-1}}{\m}-\tanh\frac{x_0}{z_H}}\)}{\sqrt{\m^2-1}}\], \hspace{4em}
\eea
where we have adopted the following conventions
\bea
&& x(\tau)+x_0=z_H\cosh^{-1}\(\m\sqrt{1-\frac{\e^2}{z_H^2}}\),\\
&&\coth\frac{x_0}{z_H}=\mathcal{G},\quad \m=\cosh\frac{x_0}{z_H}\coth\b\cos\frac{\tau}{z_H}.
\eea
Following our previous discussion, we find that in the leading order of $\eta$, the on-shell action is 
\be
I_{grav}=-I_{gravity,W}\approx-2h\log\frac{\e^2}{2z_H\sinh\frac{w_{12}}{2z_H}\times2z_H\sinh\frac{\bar{w}_{12}}{2z_H}}
\ee
as desired.

\section{Backreacted geometries}
\label{sec:backreacted geometries}
\renewcommand{\theequation}{4.\arabic{equation}}
\setcounter{equation}{0}
Instead of employing the wedge construction, in AdS$_3$ we can directly construct the backrecated geometries by following a similar approach proposed in \cite{Nozaki:2013wia,Caputa:2014eta}, which will be explained below. We begin with the metric \eqref{conical AdS3} of the conical AdS, \be \label{conical}
ds^2=(r^2+\alpha^2)d\tau^2+\frac{dr^2}{r^2+\alpha^2}+r^2d\theta^2,
\ee 
which we recall here for convenience.
To construct the backreacted geometry corresponding to a two-point function in the \Poincare AdS, we perform the following coordinate transformation
\bea\label{trpoincare}
\sqrt{r^2+1}\cosh\tau &=& \cosh\b \frac{z^2+x_1^2+x_0^2+1}{2z}-\sinh\b \frac{x_0}{z}\nonumber\\
r\cos\q &=& -\sinh\b \frac{z^2+x_1^2+x_0^2+1}{2z}+\cosh\b \frac{x_0}{z}\nonumber\\
r\sin\q  &=& \frac{x_1}{z} \nonumber\\
\sqrt{r^2+1}\sinh\tau &=&\frac{z^2+x_1^2+x_0^2-1}{2z},
\eea 
which is the same as the one-parameter transformations introduced in \cite{Nozaki:2013wia, Caputa:2014eta}, 	which facilitate the construction of a holographic dual description for a local quench model up to a shift in the $x_0$ direction. The resulting geometry is very complicated, corresponding to a \Poincare spacetime deformed by the backreaction of a local excitation. We will refer to this backreacted geometry as the ``deformed geometry".
It is important to emphasize that we previously used these transformations \eqref{boost poincare} in the wedge construction, but here we use them differently. In the earlier case, the coordinates $(r,\tau,\theta)$ described the global AdS$_3$ with metric \eqref{global_metric} and these transformations mapped the global AdS spacetime to the \Poincare AdS spacetime.
In the current context, the meaning of transformations \eqref{trpoincare} is more intricate than they appear. As noted in \cite{Tian:2024fmo}, these transformations represent a combination of a GtP transformation and a special conformal transformation. Moreover, while the deformed \Poincare solution derived from these transformations is very complicated, the on-shell action of this backreacted geometry can be computed quite straightforwardly. The computation reduces to a surface integral over the non-standard cut-off surface at $z=\epsilon$, involving the non-trivial determinant of the induced metric:
\bea\label{deformed poincare}
\sqrt{h}=\frac{1}{\e^2}-\frac{2\(1-\a^2\)}{\sinh^{2}\b\(x_1^2+\(x_0-\tanh\frac{\b}{2}\)^2\)\(x_1^2+\(x_0-\coth\frac{\b}{2}\)^2\)}.
\eea  
Similarly, other backreacted geometries can be constructed by applying the coordinate transformations \eqref{boost global} and \eqref{boost btz}.

\bigskip

Before proceeding with the calculation of the on-shell action in this background geometry, we outline here also the essential geometric setup required for the calculation of the on-shell action in the FG gauge for future use, which takes the general form
\be 
ds^2_{\rm FG}=\frac{dz_{f}^2+\(du_{f}+z_f^2\bar{\mathcal{L}}(v_{f}) dv_f\)\(dv_{f}+z_f^2\mathcal{L}(u_{f})du_{f}\)}{z_{f}^2}.
\ee 
Different solutions are characterized by the functions $\mathcal{L}(u_f)$ and $\bar{\mathcal{L}}(v_f)$, which are related to the expectation value of the energy-stress tensor via \cite{Banados:1998gg}:
\be 
\mathcal{L}(u_f)=-\frac{6}{c}\langle T(u_f)\rangle,\quad \bar{\mathcal{L}}(v_f)=- \frac{6}{c}\langle \bar{T}(v_f)\rangle.
\ee 
For instance, for the conical AdS geometry, the corresponding functions are
\be \label{lt}
\mathcal{L}=\frac{\alpha^2}{4},\quad \mathcal{\bar{L}}=\frac{\alpha^2}{4}.
\ee
To derive $\mathcal{L}(\bar{\mathcal{L}})$ for other deformed geometries, one can either solve the FG gauge directly by making an ansatz in the form of an expansion with respect to the AdS radial coordinate \cite{Nozaki:2013wia} or use the transformation rule of the energy-stress tensor
\bea\label{Ttbar}
\braket{T(u)}=\(\frac{dw}{du}\)^{2}\braket{T(w)}_{}+\frac{c}{12}\text{Sch}\(w,u\)
\eea
where the Schwarzian derivative is defined as
\be
\text{Sch}(w,u)=\frac{w'''(u)}{w'(u)}-\frac{3\(w''(u)\)^2}{2\(w'(u)\)^2},
\ee
with $u,w$ being the light-cone coordinates at the AdS boundary. The second method can also tell us the relationship between the CFT states that are dual to the bulk solutions before and after transformations. 
For example, as we approach the AdS boundary, the transformations \eqref{trpoincare} reduce to
\bea \label{conformal1}
u= \frac{\sinh\frac{\beta}{2}+e^{ w}\cosh\frac{\beta}{2}}{\cosh\frac{\beta}{2}+e^{w}\sinh\frac{\beta}{2}},\quad v= \frac{\sinh\frac{\beta}{2}+e^{\bar{w}}\cosh\frac{\beta}{2}}{\cosh\frac{\beta}{2}+e^{ \bar{w}}\sinh\frac{\beta}{2}},
\eea 
where the conventions of the light-cone coordinates are
\be 
w=\tau+\im \theta,\quad u=x_0+\im x_1.
\ee 
Substituting into \eqref{Ttbar} and \eqref{lt} yields 
\be 
\mathcal{L}(u)=-\frac{1-\alpha^2}{4}\frac{u_{ab}^2}{(u-u_a)^2(u-u_b)^2},
\ee 
where
\be 
u_a\equiv \coth\frac{\b}{2},\quad u_b\equiv\tanh\frac{\b}{2}.
\ee 
Notably, since $\cosh^2(\frac{\beta}{2})-\sinh^2 (\frac{\beta}{2})=1$, the transformation \eqref{conformal1} can be interpreted as a compound transformation $u=f_\beta(e^w)$, where $f_\beta$ is a \text{M$\ddot{\text{o}}$bius} transformation. Recall that conical AdS geometry is dual to a primary state; thus the transformation \eqref{conformal1} implies that the CFT state corresponding to the deformed \Poincare geometry is given by
\be
|h\rangle\xlongrightarrow{e^{w}}\mathcal{O}(0)|\Omega\rangle\xlongrightarrow{\text{M$\ddot{\text{o}}$bius}}\mathcal{O}(u_a)|\Omega\rangle,\quad u_a=\tanh\frac{\beta}{2}.
\ee

Similarly, we find that FG gauges of the deformed global AdS and deformed BTZ are characterized by\footnote{In the deformed BTZ, in order to obtain the correct result, the light-cone coordinate should be chosen as $u_{\rm BTZ}=x+\im\tau_{\rm BTZ}$ due to the non-standard transformation we have chosen.} 
\bea 
&&\mathcal{L}(u_{\text{global}})=\frac{1}{4}-\frac{1-\alpha^2}{4(\cosh\beta-\sinh\beta \cosh u_{\text{global}})^2},\label{global:FG}\\
&&\mathcal{L}(u_{\text{BTZ}})=-\frac{1}{4z_H^2}-\frac{1-\alpha^2}{4z_H^2(\cosh\beta\cosh(u_{\text{BTZ}}/z_H)-\sinh\beta)^2}.\label{btz:FG}
\eea 

\subsection{The on-shell action}
After outlining our general approach, we present the calculation of the on-shell actions for the deformed geometries derived from the transformation, as well as those in their FG gauge. Specifically, we compute the gravitational action defined by \eqref{action_gravity}, under the assumption that the matter term is canceled by the gravitational contribution associated with the conical singularity. We will demonstrate that the on-shell actions of the former are directly related to the correlation functions, whereas the on-shell actions of the latter are not. This distinction represents one of the key results of this paper.

\subsection*{On-shell action of deformed \Poincare AdS}
We begin with the deformed \Poincare AdS geometry. As mentioned above, we apply the above transformation \eqref{trpoincare} into conical AdS$_3$, it's straightforward to show the determinant is invariant under the transformation
\bea
\sqrt{g}=\frac{1}{z^3}
\eea
and the determinant of the induced metric on a naive cutoff surface is given by \eqref{deformed poincare}. So the on-shell action in this deformed \Poincare patch is given by
\bea
I^{BR}&=&-\frac{1}{16\pi G_N}\int_{\cM}\sqrt{g}\(R+2\)-\frac{1}{8\pi G_N}\int_{\pcM}\sqrt{h}\(K-1\)+m\int_{\G}\sqrt{\g}\nn
&=&\frac{1-\a^2}{4\pi G_N}\frac{\left|u_{ab}\right|^2}{4}\int_{\pcM}\frac{\sqrt{g^{(0)}}du dv}{\left|u-u_a\right|^2\left|u-u_b\right|^2}\nonumber\\
&=&\frac{h \left|u_{ab}\right|^2}{\pi}\int_{\pcM}\frac{\sqrt{g^{(0)}}du dv}{\left|u-u_a\right|^2\left|u-u_b\right|^2}
\eea
where the boundary metric
\bea
ds^2=g^{\(0\)}_{ij}dx^{i} dx^{j}=du dv
\eea
have been introduced. The detail of this integral can be found in the appendix \ref{append:B}. We present the final result as follows
\be
I^{BR}=\frac{h \left|u_{ab}\right|^2}{\pi}\times\frac{2\pi }{\left|u_{ab}\right|^2}\log\frac{u_{ab}v_{ab}}{\e^2}=4h\log\frac{\left|u_{ab}\right|}{\e}
\ee
which indicates the partition function defined in the constructed solution exactly reproduces the two-point correlation function
\be
e^{-I^{BR}}=\frac{\e^{2\(h+\bar{h}\)}}{\(u_{ab}\)^{2h}\(v_{ab}\)^{2\bar{h}}}=\(\frac{\e}{\left|u_{ab}\right|}\)^{2\Delta}.
\ee
where $h=\bar{h}=\frac{\D}{2}$ for scalar primary operators. 
 
\begin{figure}[htbp]
	\centering
	\includegraphics[width=0.45\textwidth,scale=2]{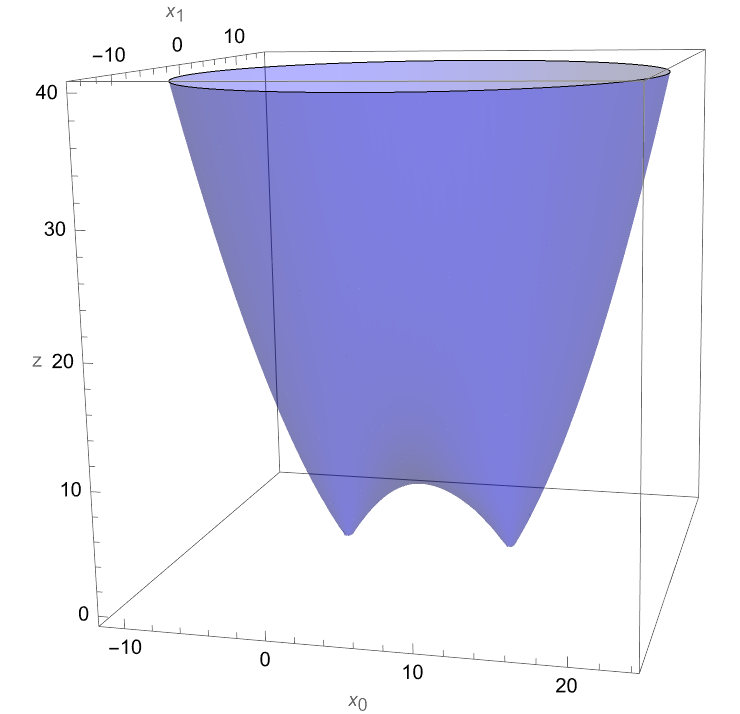}
	\caption{Image of the Wall in deformed \Poincare AdS.}
	\label{fig:imageofwall}
\end{figure}

While in its FG patch, where a wall at which the determinant of the FG expansion becomes degenerate always exists in the bulk and its expression is determined by
\bea\label{FG:wall}
z_{f}=z_{w},\quad z_{w}^{-4}=\mathcal{L}(u_{f})\bar{\mathcal{L}}(v_{f}).
\eea
The on-shell action takes the following form
\bea\label{FG:integral}
I^{BR}_{\rm FG}&=&\frac{c}{6\pi}\int_{\pcM}\sqrt{g^{(0)}}du_{f} dv_{f}\int_{\e}^{z_w}\(\frac{1}{z_{f}^3}-\mathcal{L}(u_{f})\bar{\mathcal{L}}(v_{f}) z_{f}\)dz_{f}-\frac{c}{12\pi}\int_{\pcM}\sqrt{g^{(0)}}du_{f} dv_{f}\,\frac{1}{\e^2}\nonumber\\
&=&-\frac{c}{6\pi}\int_{\pcM}\sqrt{g^{(0)}}du_{f} dv_{f}\,\sqrt{\mathcal{L}(u_{f})\bar{\mathcal{L}}(v_{f})}
\eea
in general. In this case, substituting the \eqref{deformed poincare} into above integral gives
\bea
I^{BR}_{\rm FG}=-2h\log\frac{u_{ab}v_{ab}}{\e^2}
\eea
which implies the partition function in FG patch gives the inverse of the two-point function. This can be interpreted as indicating that the FG coordinates cover only the region outside the following surface in deformed \Poincare AdS, given by
\be
z^{-2}=\frac{u_{ab}v_{ab}\(1-\a^2\)}{\(u-u_a\)\(u-u_b\)\(v-v_a\)\(v-v_b\)}
\ee 
which is the image of the wall, as showed in Figure \ref{fig:imageofwall}. When $\b\rightarrow 0$, this surface reduces to a conical surface $\frac{R}{z}=\sqrt{1-\a^2}$ as presented in \cite{Tian:2024fmo}.  
It is straightforward to verify that the on-shell action computed in the region outside the image surface of the wall is identical to that in the FG coordinates:
\bea
I_{image, out}^{BR}&=&I^{BR}_{\rm FG}
\eea

\subsection*{On-shell action of deformed global AdS}
In deformed global AdS, the determinant is also invariant under the transformation
\be
\sqrt{g}=r
\ee
and the determinant of the induced metric at a naive cutoff $r=\frac{1}{\e}$ is given by
\bea
\sqrt{h}&=&\frac{1}{\e^2}+\frac{1}{2}\(1-\frac{1-\a^2}{\sinh^2\b\(\cosh w-\coth\b\)\(\cosh \bar{w}-\coth\b\)}\)\nonumber\\
&=&\frac{1}{\e^2}+\frac{1}{2}\(1-\frac{1-\a^2}{4}\frac{\sinh w_{b}\sinh\bar{w}_b}{\sinh\frac{w-w_b}{2}\sinh\frac{w+w_b}{2}\sinh\frac{\bar{w}-\bar{w}_b}{2}\sinh\frac{\bar{w}+\bar{w}_b}{2}}\)
\eea
where
\be
\cosh w_b\equiv\coth\b, \quad w_b=\tau_b.
\ee
In this case, the on-shell action defined in deformed global AdS is 
\bea
I^{BR}&=&-\frac{1}{16\pi G_N}\int_{\cM}\sqrt{g}\(R+2\)-\frac{1}{8\pi G_N}\int_{\pcM}\sqrt{h}\(K-1\)+m\int_{\G}\sqrt{\g}\nn
&=&-\frac{1}{16\pi G_N}\int_{\pcM}dw d\bar{w}\sqrt{g^{(0)}}\(1-\frac{1-\a^2}{4}\frac{\sinh w_{b}\sinh\bar{w}_b}{\sinh\frac{w-w_b}{2}\sinh\frac{w+w_b}{2}\sinh\frac{\bar{w}-\bar{w}_b}{2}\sinh\frac{\bar{w}+\bar{w}_b}{2}}\)\nn
&=&-\frac{W T}{16\pi G_N}+\frac{h \sinh w_{b}\sinh\bar{w}_b}{4\pi}\int_{\pcM}\frac{\sqrt{g^{(0)}}dw d\bar{w}}{\left|\sinh\frac{w-w_b}{2}\right|^2\left|\sinh\frac{w+w_b}{2}\right|^2}\nonumber\\
&=&I_{\rm global,vacuum}+2h\log\frac{2\sinh w_b\times 2\sinh \bar{w}_b}{\e^2}\label{action:global}
\eea 
where $W=2\pi$ represents the width of the boundary cylinder, $T$ is the length of the boundary cylinder or the period along the Euclidean time $\tau$ direction. In the final step, we performed the transformation $u=e^{w}$, which maps an infinite cylinder onto the complex plane. This allows the integral to be evaluated as follows:
\bea\label{global:integral}
\int_{\pcM}\frac{\sqrt{g^{(0)}}dw d\bar{w}}{\left|\sinh\frac{w-w_b}{2}\right|^2\left|\sinh\frac{w+w_b}{2}\right|^2}=16\left|u_{1}\right|\left|u_{2}\right|\int_{\pcM} \frac{\sqrt{g^{(0)}}du dv}{\left|u-u_1\right|^2\left|u-u_2\right|^2}=\frac{32\pi\left|u_1\right|\left|u_2\right|}{\left|u_{12}\right|^2}\log\frac{u_{12}v_{12}}{\(2\delta\)^{2}}.\quad 
\eea
Recall that
\be
\frac{4\left|u_1\right|\left|u_2\right|}{\left|u_{12}\right|^2}=\frac{1}{\sinh\frac{w_{12}}{2}\sinh\frac{\bar{w}_{12}}{2}}=\frac{1}{\sinh w_{b}\sinh\bar{w}_{b}}, \quad w_{1,2}=\pm\t_b
\ee
up to a redefinition of the cutoff
\be
\frac{1}{\e^2}=\frac{\left|u_{1}\right|\left|u_{2}\right|}{4\delta^2}=\frac{1}{4\delta^2}, \quad u_{1,2}=e^{\mp\tau_b},
\ee
the above result is successfully reproduced, yielding the partition function as
\bea
\frac{e^{-I_{BR}}}{e^{-I_{\rm global,vacuum}}}=\frac{\e^{2h+2\bar{h}}}{\(2\sinh\frac{w_{12}}{2}\)^{2h} \(2\sinh\frac{\bar{w}_{12}}{2}\)^{2\bar{h}}}=\(\frac{\e^2}{2\sinh\frac{w_{12}}{2}\times 2\sinh\frac{w_{12}}{2}}\)^{\Delta}
\eea
with $h=\bar{h}=\frac{\Delta}{2}$ for scalar primary operators. The result matches the two-point function in global AdS up to an analytic continuation $\tau=\im t$, which arises from the different convention adopted here.

It is important to emphasize that the calculation above relies on the implicit assumption that the period $T$ along the Euclidean time $\tau$ direction is either non-compact or much larger than the insertion points, $T\gg 2\tau_b$.  This assumption allows us to neglect the boundary terms appearing in the integral \eqref{global:integral}, as detailed in Appendix \ref{append:B}. Specifically, if we assume the range of $\tau$ is $-\frac{T}{2}<\tau<\frac{T}{2}$, then the boundary cylinder is mapped to an annulus $e^{-\frac{T}{2}}<R< e^{\frac{T}{2}}$ in the complex plane. In evaluating the integral, we should include the following boundary term\footnote{In global AdS and BTZ spacetimes, we do not impose an identification between the two edges; hence, the boundary geometry is not a torus.}
\be
2h\log\frac{\cosh T-\cosh(2\tau_b)}{\cosh T-1},
\ee
which is of constant order. Obviously, when $T\gg 2\t_b$, its contribution can be neglected.

However, in the FG gauge, the on-shell action is no longer the negative of that \eqref{action:global} in deformed global AdS, as shown below. In the FG coordinate, a general function $\mathcal{L}(\bar{\mathcal{L}})$  can be obtained by using Conformal Ward identity, and the result is
\be
\mathcal{L}(w_f)=\frac{1}{4}\(1-\frac{6h}{c}\frac{\sinh^2\frac{w_{12}}{2}}{\sinh^2\frac{w_f-w_{1}}{2}\sinh^2\frac{w_f-w_{2}}{2}}\)
\ee
which is consistent with \eqref{global:FG} once the insertion points are substituted. Substituting this function into the integral \eqref{FG:integral} gives the on-shell action in FG coordinate
\bea
I_{\rm FG}^{BR}=-\frac{c}{24\pi}\int_{\pcM}d^{2}w_f\sqrt{g^{(0)}}f_{\rm FG}(w_{f},\bar{w}_f)  
\eea
where 
\be
f_{\rm FG}= 4\(\mathcal{L}\bar{\mathcal{L}}\)^{\frac{1}{2}}=\left|1-\frac{\(1-\alpha^2\)\csch^2\b}{(\coth\beta-\cosh w_f)^2}\right|
\ee
To verify whether $I_{\rm FG}^{BR}$ yields the inverse of the two-point correlation function, i.e,
\bea
\frac{e^{-I_{\rm FG}^{BR}}}{e^{-I_{\rm global,vacuum}}}\xlongequal{?}\(\frac{\e^2}{2\sinh\frac{w_{12}}{2}\times 2\sinh\frac{\bar{w}_{12}}{2}}\)^{-2h},
\eea
we need to know the result of the above integral. However, it is challenging to evaluate this integral analytically and verify the equality, although some numerical evidence and perturbative results suggest that this equality does not hold. In order to provide a more definitive answer, here we take the other way around as follows.

Let us first revisit the on-shell action in FG coordinates. For a generic upper boundary $z_f(w_f,\bar{w}_{f})$, the action is given by
\bea
I_{\rm FG}=-\frac{c}{12\pi}\int_{\pcM}d^{2}w\sqrt{g^{(0)}}\(z_f^{-2}+z_f^2\mathcal{L}\bar{\mathcal{L}}\)
\eea
in which the integrand is a monotonic function of $z$ since the range of $z$ is limited to $0<z_f^2<\(\mathcal{L}\bar{\mathcal{L}}\)^{-\frac{1}{2}}=z_{w}^{2}, z_{f}\geq 0$.  Now assuming the action $I_{\rm FG}$ exactly gives the inverse of the two-point function, then it's straightforward to show the upper boundary surface is determined by
\bea
z_f^2 \mathcal{L}(w_{f})\bar{\mathcal{L}}(\bar{w}_f)+z_f^{-2}=\frac{1+\frac{\(1-\a^2\)}{4}\frac{\left|\sinh\frac{w_{12}}{2}\right|^2}{\left|\sinh\frac{w_f-w_1}{2}\sinh\frac{w_f-w_2}{2}\right|^2}}{2}.
\eea
A key observation is
\bea
\(\frac{1+\frac{\(1-\a^2\)}{4}\frac{\left|\sinh\frac{w_{12}}{2}\right|^2}{\left|\sinh\frac{w_f-w_1}{2}\sinh\frac{w_f-w_2}{2}\right|^2}}{2}\)^2-4\mathcal{L}(w_{f})\bar{\mathcal{L}}(\bar{w}_f)=\frac{1}{4}\(\frac{\sinh\frac{w_{f,12}}{2}}{\sinh\frac{w_{f}-w_1}{2}\sinh\frac{w_{f}-w_2}{2}}+\rm{Conjugate}\)^2\quad 
\eea
which is always non-negative. This observation implies that for a given boundary coordinate, the point  $z_{f}(w_{f},\bar{w}_f)$ on this surface is lower than that in the wall
\bea
z_{f}^2\leq z_{w}^2=\(\mathcal{L}\bar{\mathcal{L}}\)^{-\frac{1}{2}}.
\eea
The result is  the surface, under which the on-shell action yields the inverse of the correlation function, lies entirely outside the wall except at the singularity. Besides, the action between the wall and this surface is nonzero. Supporting evidence arises in the limit $\b\rightarrow 0$, where the surface $z_{f}\rightarrow \frac{2}{1+\sqrt{1-\a^2}}$ and $z_{w}\rightarrow \frac{2}{\a}$. Clearly the surface $z_{f}$ is under the wall. In this limit, the on-shell action outside the the surface $z_f(w_{f},\bar{w}_f)$ is given by
\bea
I_{\rm FG, \b=0}&=&-\frac{c}{12\pi}\int_{\pcM}d^{2}w\sqrt{g^{(0)}}\(\frac{1}{2}+\frac{1-\a^2}{2}\)=I_{\rm global,vacuum}-\frac{h}{\pi} W T\nn
&=& I_{\rm global,vacuum}-2h\log\frac{\L}{\e}, \quad W=2\pi
\eea
which is smaller than $I^{BR}_{\rm FG,\b=0}=-\frac{c\a^2}{24\pi}WT$. Here, the identification between the Euclidean time direction and the radial direction in the complex plane leads to $T=\log\frac{\Lambda}{\e}$ \cite{Tian:2024fmo}. Clearly, the inverse of the one-point function can then be read from the partition function
\be
\frac{e^{-I_{\rm FG,\b=0}}}{e^{-I_{\rm global,vacuum}}}=\(\frac{\L}{\e}\)^{\D},
\ee
which is consistent with the above discussion. Therefore, $I_{\rm FG}^{BR}$ cannot give the inverse of the two-point function in this case, and 
\bea
I_{\rm FG}^{BR}>I_{\rm global,vacuum}-2h\log \frac{2\sinh\frac{w_{12}}{2}\times 2\sinh \frac{\bar{w}_{12}}{2}}{\e^2}
\eea
In conclusion, the partition function in the FG coordinate is exactly smaller than the inverse of the correlation function 
\bea
\frac{e^{-I^{BR}_{\rm FG}}}{e^{-I_{\rm global,vacuum}}}<\(\frac{\e^2}{2\sinh\frac{w_{12}}{2}\times 2\sinh\frac{\bar{w}_{12}}{2}}\)^{-2h}.
\eea

\subsection*{On-shell action of deformed BTZ black hole}
In deformed BTZ black hole, the determinant of the metric
\bea
\sqrt{g}&=&\frac{1}{z^3}
\eea
is invariant as expected, and the induced metric defined at a naive cutoff $z=\e$ is given by
\bea
 \sqrt{h} &=&\frac{1}{\e^2}-\frac{1}{2z_H^2}\(1+\frac{1-\a^2}{\cosh^2\b}\frac{1}{\left|\cosh\frac{w}{z_H}-\tanh\b\right|^2}\)\nn &=&\frac{1}{\e^2}-\frac{1}{2z_H^2}\(1+\frac{1-\a^2}{4}\frac{\left|\sinh\frac{w_{12}}{2z_H}\right|^2}{\left|\sinh\frac{w-w_1}{2z_H}\right|^2\left|\sinh\frac{w-w_2}{2z_H}\right|^2}\)
\eea
where the light-cone coordinate is defined as follows
\be
w=x+\im\tau, \quad w_{1,2}=\pm\im z_H\cos^{-1}(\tanh\b).
\ee
The on-shell action defined in this constructed solution is 
\bea
I^{BR}&=&-\frac{1}{16\pi G_N}\int_{\cM}\sqrt{g}\(R+2\)-\frac{1}{8\pi G_N}\int_{\pcM}\sqrt{h}\(K-1\)+m\int_{\G}\sqrt{\g}\nn
&=&-\frac{WT}{16\pi G_N z_H^2}+\frac{h\left|\sinh\frac{w_{12}}{2z_H}\right|^2}{4\pi z_H^2}\int_{\pcM}\frac{\sqrt{g^{(0)}}d^2w}{\left|\sinh\frac{w-w_1}{2z_H}\right|^2\left|\sinh\frac{w-w_2}{2z_H}\right|^2}\nonumber\\
&=&I_{\rm BTZ,vacuum}+2h\log\frac{2z_H\sinh\frac{w_{12}}{2z_H}\times2z_H\sinh\frac{\bar{w}_{12}}{2z_H}}{\e^2}+I_{bdy}.
\eea
 Here, $W$ and $T=2\pi z_H$ represent the width and the length of boundary cylinder of the deformed BTZ black hole, respectively. In the final step, we applied the transformation $u=e^{w/z_H}$ and the result is
\bea
\int_{\pcM}\frac{\sqrt{g^{(0)}}dw d\bar{w}}{\left|\sinh\frac{w-w_1}{2z_H}\right|^2\left|\sinh\frac{w-w_2}{2z_H}\right|^2}=\int_{\pcM}\frac{16z_H^2\left|u_{1}\right|\left|u_{2}\right|\sqrt{g^{(0)}}du dv}{\left|u-u_{1}\right|^2\left|u-u_{2}\right|^2}= \frac{32\pi z_H^2\left|u_{1}\right|\left|u_{2}\right|}{\left|u_{12}\right|^2}\log\frac{\left|u_{12}\right|^2}{\(2\delta\)^2}.\quad 
\eea
With 
\be
\frac{\left|u_{1}\right|\left|u_{2}\right|}{\left|u_{12}\right|^2}=\frac{1}{4\sinh\frac{w_{12}}{2z_{H}}\sinh\frac{\bar{w}_{12}}{2z_{H}}},
\ee
up to an identification
\be
\frac{1}{\e^2}=\frac{\left|u_{1}\right|\left|u_{2}\right|}{4z_H^2 \delta^2},
\ee
we obtain the result presented above. The discussion about the boundary term follows an analogy to that in global AdS. However, due to the specific definition of the light-cone coordinate in this context, the two insertions are mapped to the unit circle with a phase difference of $\pi$. If the spatial direction is non-compact, then the boundary term $I_{bdy}$ vanishes. Otherwise, the boundary term contributes at a constant order as follows
\bea
I_{bdy}=2h\log\frac{\cosh W-1}{\cosh W+1-2\tanh^2\b}.
\eea
where the range of $x$ is set to $-\frac{W}{2} <x<\frac{W}{2}$. So the leading order partition function gives
\be
\frac{e^{-I_{BR}}}{e^{-I_{\rm BTZ,vacuum}}}\approx\(\frac{\e^2}{2z_H\sinh\frac{w_{12}}{2z_H}\times2z_H\sinh\frac{\bar{w}_{12}}{2z_H}}\)^{\Delta},
\ee
which is exactly the two-point function in BTZ blackhole. 

In the FG coordinate of the deformed BTZ black hole, substituting the $\mathcal{L}(\bar{\mathcal{L}})$ function \eqref{btz:FG} into the expression \eqref{FG:wall} of the wall  and  the integral \eqref{FG:integral} yields the location of the wall and the on-shell action $I_{\rm FG}^{BR}$, respectively, in this case. As previously discussed, it is challenging to directly verify whether $I_{\rm FG}^{BR}$ gives the inverse of the two-point function 
\bea
\frac{e^{-I_{\rm FG}^{BR}}}{e^{-I_{\rm BTZ,vacuum}}}\xlongequal{?}\(\frac{\e^2}{2z_H\sinh\frac{w_{12}}{2}\times 2z_H\sinh \frac{\bar{w}_{12}}{2}}\)^{-2h}.
\eea
Therefore, here we take the other way around as in previous case. The surface, under which the on-shell action in FG coordinates yields the inverse of the correlation function, is determined by
\bea
z_f^2 \mathcal{L}(w_{f})\bar{\mathcal{L}}(\bar{w}_f)+z_f^{-2}=\frac{1}{2z_H^2}\(1+\frac{1-\a^2}{4}\frac{\left|\sinh\frac{w_{12}}{2z_H}\right|^2}{\left|\sinh\frac{w-w_1}{2z_H}\right|^2\left|\sinh\frac{w-w_2}{2z_H}\right|^2}\).
\eea
Notice that
\bea
\frac{\(1+\frac{1-\a^2}{4}\frac{\left|\sinh\frac{w_{f,12}}{2z_H}\right|^2}{\left|\sinh\frac{w_f-w_1}{2z_H}\right|^2\left|\sinh\frac{w_f-w_2}{2z_H}\right|^2}\)^2}{4z_H^4}-4\mathcal{L}\bar{\mathcal{L}}=-\frac{1-\a^2}{4}\(\frac{\sinh\frac{w_{f,12}}{2z_H}}{\sinh\frac{w_f-w_{1}}{2z_H}\sinh\frac{w_f-w_{2}}{2z_H}}-\text{Conjugate}\)^2\;\;
\eea
which is strictly non-negative. As in global AdS, the result implies that the surface, under which the on-shell action reproduces the inverse of the two-point function, lies entirely outside the wall,  except at the singularity. When $\beta\rightarrow 0$, this surface reduces to
\bea
z_f^{-2}+\mathcal{L}\bar{\mathcal{L}}z_f^2=\frac{1}{2z_H^2}+\frac{1-\a^2}{2z_H^2}\frac{1}{\cosh\frac{w_f}{z_H}\cosh\frac{\bar{w}_f}{z_H}}.
\eea
In this limit, the on-shell action evaluated under this surface is given by
\bea
I_{\rm FG,\b=0}&=&-\frac{c}{12\pi}\int_{\pcM}d^{2}w\sqrt{g^{(0)}}\(\frac{1}{2z_H^2}+\frac{1-\a^2}{2z_H^2}\frac{1}{\cosh\frac{w}{z_H}\cosh\frac{\bar{w}}{z_H}}\)\nn
&=&-I_{\rm BTZ,vacuum}-2h\log\frac{2\sinh\frac{w_{12}}{2z_H}\times 2\sinh\frac{\bar{w}_{12}}{2z_H}}{\e^2}
\eea
where the boundary insertions reduce to $w_{1,2}=\im\tau_{1,2}=\pm\im\frac{\pi z_H}{2}$ in this case. Therefore, the partition function $e^{-I_{\rm FG,\b=0}}$ in the FG gauge gives the inverse of the two-point function.

In conclusion, the on-shell action $I_{\rm FG}^{BR}$ defined in the FG patch does not yield the inverse of the two-point function in this case. Moreover, the former is generally larger than the latter, as demonstrated below
\bea
I_{\rm FG}^{BR}>I_{\rm BTZ,vacuum}-2h\log \frac{2z_H\sinh\frac{w_{12}}{2z_H}\times 2z_H\sinh\frac{w_{12}}{2z_H}}{\e^2},
\eea
which strengthen our belief that the action inside the FG patch is not always meaningful \cite{Tian:2024fmo}. We leave this interesting question to the future.

\section{Discussion}
\label{sec:discussion}
In this work, we proposed and examined a novel holographic approach for computing the correlation functions of operators in CFTs, refining and generalizing the proposal presented in \cite{Caputa:2022zsr}. We demonstrated that correlation functions can be computed through the on-shell actions of excised geometries in the probe limit by doing numeric and perturbative calculations. These geometries are constructed from various AdS solutions, including \Poincare AdS, global AdS, and BTZ solutions, by excising a wedge bounded by two intersecting EOW branes and the AdS asymptotic boundary. From the perspective of the AdS/BCFT correspondence, this wedge is dual to a BCFT defined in a region with boundary cusps\footnote{Interestingly, it is recently pointed out in \cite{Miao:2024ddp} that the BCFT defined within a wedge region with non-smooth corners may exhibit certain singularities in the energy-stress tensor.}. 

Additionally, we constructed the backreacted geometries corresponding to the correlation functions using a more direct approach. This involved performing coordinate transformations on the conical geometry. We computed the on-shell actions for these backreacted solutions, as well as for their FG gauges.  We found that the on-shell actions of the backreacted solutions are proper to reproduce the correlation functions with no need of taking the probe limit, while they differ from the on-shell actions of the same backreacted geometry in FG gauge. This discrepancy, previously noted in \cite{Tian:2024fmo}, is supported by additional examples in this work. The AdS$_3$ solutions in the FG gauges, known as the \Banados solutions, are widely utilized in AdS$_3$ holography. Our findings may reveal deeper insights that have been previously overlooked. A potential approach to addressing this issue involves employing a generalized FG gauge, as proposed in \cite{Arenas-Henriquez:2024ypo}. We discuss several future directions below.

\textbf{Local Quench.} As mentioned in Section \ref{sec:backreacted geometries}, the backreacted (deformed) solutions can serve as models for studying local quenches. It would be intriguing to investigate other properties, such as entanglement or entwinement \cite{Balasubramanian:2014sra}, within these excised geometries.

\textbf{Correlation Functions of Boundary Condition Changing (BCC) Operators.} In this work, we focused on correlation functions in a CFT. However, when motivating our wedge proposal, we also noted that the generalized AdS/BCFT correspondence allows us to study BCC operators. We proposed a straightforward, albeit perhaps naive, setup for computing the two-point functions of BCC operators. Further exploration in this area would be valuable.

\textbf{Spinning Operators.} Our study concentrated on the correlation functions of scalar operators. Recent work has highlighted interesting developments regarding spinning particles in AdS$_3$ \cite{mathisson1937neue,papapetrou1951spinning,dixon1970dynamics,Castro:2014tta,Chang:2018nzm,Li:2024rma,Baake:2023gxx,Briceno:2021dpi}. In particular, the authors of \cite{Kusuki:2022ozk} proposed an excised geometry to describe spinning defects. We anticipate that our wedge proposal can also be extended to include correlation functions of spinning operators.

\textbf{Three-Dimensional C-Metric.} Recently, C-metric solutions in AdS$_3$, which describe accelerating particles or black holes, have garnered renewed attention. As shown in \cite{Astorino:2011mw,Xu:2011vp}, a 3D C-metric can be derived from a direct truncation of solutions in four dimensions, which have been explored extensively \cite{Letelier:1998rx,Bicak:1999sa,Podolsky:2000at,Pravda:2000zm,Dias:2002mi,Griffiths:2005qp,Krtous:2005ej,Dowker:1993bt,Emparan:1999fd,Emparan:1999wa,Emparan:2000fn,Gregory:2008br}. Recent studies have clarified the holography of the C-metric in three dimensions \cite{Arenas-Henriquez:2022www,Arenas-Henriquez:2023hur,Arenas-Henriquez:2023vqp,Tian:2023ine,Tian:2024mew}. As pointed out in \cite{Tian:2023ine,Tian:2024mew}, due to the EOW brane construction, the 3D C-metric is naturally dual to a CFT with defects. Therefore, we expect that a suitable excised geometry based on the 3D C-metric will be dual to correlation functions in a CFT with defects or boundaries.

\begin{figure}[hbtp]
	\centering
	\includegraphics[width=0.5\textwidth]{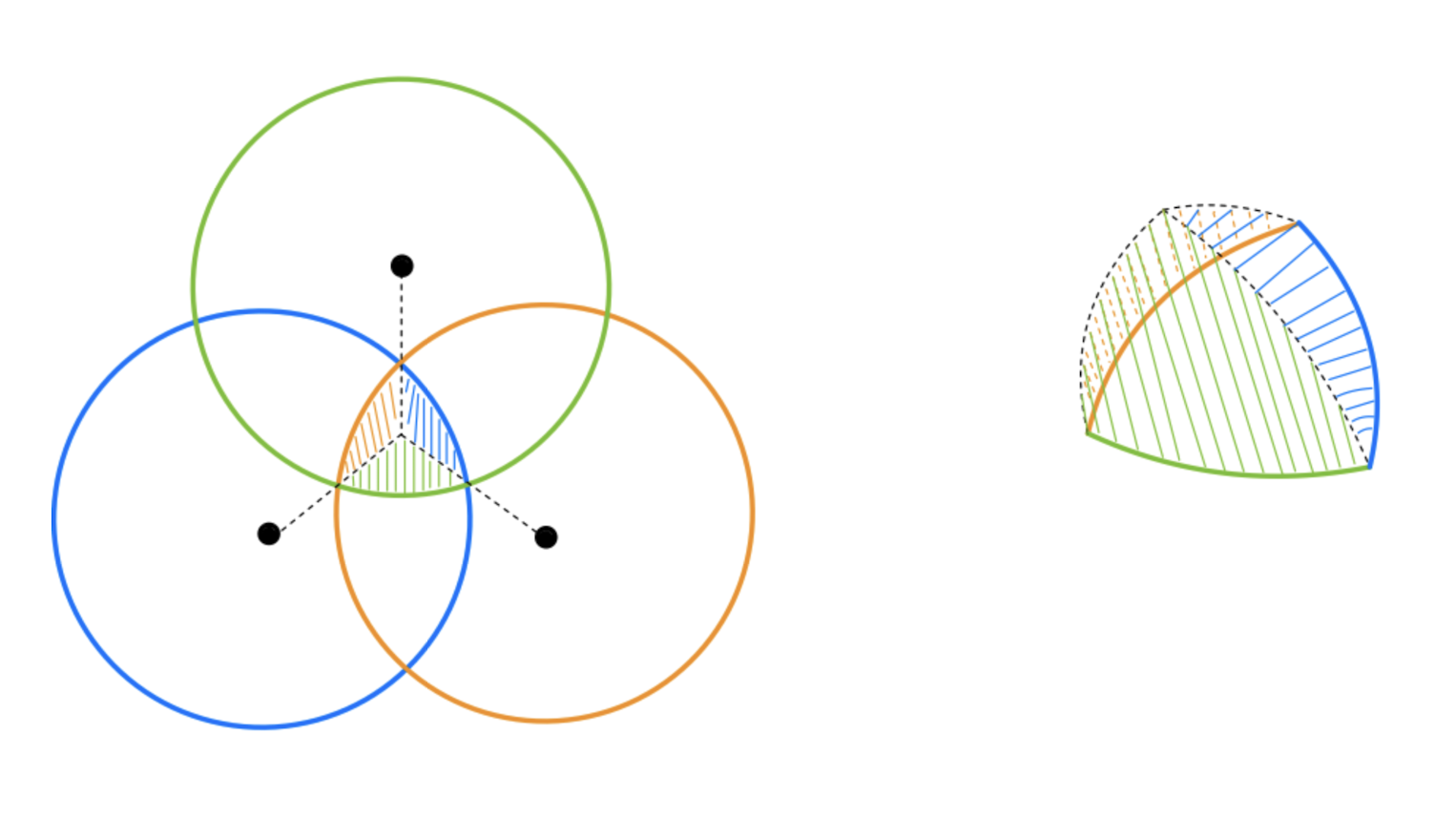}
	\caption{A naive attempt at constructing a wedge for a three-point function, represented as the overlap of three hemispheres.}
	\label{three-point}
\end{figure}

\textbf{Higher Dimensions and Higher-Point Functions.} Lastly, it would be intriguing to generalize the wedge proposal to higher dimensions, where constructing a wedge geometry is significantly more straightforward than a backreacted geometry. The extension to higher-point functions is essential yet challenging. We have verified that a naive configuration, such as the one illustrated in Figure \ref{three-point}, does not yield valid results. One possible direction is to consider certain holographic polygons \cite{Miao:2024ddp}. We will leave these explorations for future studies.
			
\section*{Acknowledgments}
We thank Bo-Hao Liu and Wen-Bin Pan for the valuable discussion on related topics. We thank Farzad Omidi for discussion and collaboration on related works. T.L would like to especially thank HH for his assistance in completing the figures in this paper. JT is supported by the National Youth Fund No.12105289 and funds from the
University of Chinese Academy of Sciences (UCAS), the Kavli Institute for Theoretical
Sciences (KITS). T.L and YWS are supported by Project  12035016 of the
National Natural Science Foundation of China.

\appendix

\section{Wedge in \Poincare AdS$_3$}
\label{append:A}
\renewcommand{\theequation}{A.\arabic{equation}}
\setcounter{equation}{0}
In \Poincare AdS$_3$, the action inside the wedge region is given by\footnote{The detail of this calculation can be found in \cite{Caputa:2022zsr}, we repeat it here for sake of completeness.}
\bea
I_{\rm AdS,EH}&=& -\frac{1}{16\pi G_N}\int_{\cM}\sqrt{g}\(R+2\)-\frac{1}{8\pi G_N}\int_{\pcM}\sqrt{h}\(K-1\)\nn
&=&\frac{1}{8\pi G_N}\int_{-\h}^{+\h} d\q\int_{a^2/\cos^2(\q)}^{\sqrt{b^2-\e^2}} \frac{d\r^2}{\r^2-b^2}\nn
&=&\frac{1}{8\pi G_N}\int_{-\h}^{\h}\(2\log\frac{\e}{b}-\log\[1-\cos^2(\h)\sec^2(\q)\]\)d\q
\eea 
where the following polar coordinate 
\be
x_0=\coth\b+\r\sin\q,\quad x_1=-\frac{1}{\sinh\b \tan\h}+\rho\cos\q
\ee
and the  convention 
\bea
b=\frac{1}{\sin(\h)\sinh\b},\quad a=\frac{1}{\tan(\h)\sinh\b}=b\cos(\h)
\eea
have been adopted. Given the following integral identity
\bea
\int \log(1-\cos^2(\h)\sec^2(\q)) d\q=\frac{\im}{2}\[\text{Li}_{2}(e^{2\im\(\q-\h\)})+\text{Li}_{2}(e^{2\im\(\q+\h\)})-2\text{Li}_{2}(-e^{2\im \q})\]
\eea
and the property of dilogarithm function $\text{Li}_{2}(z)$
\bea
\text{Li}_{2}(z)+\text{Li}_{2}(-z)=\frac{1}{2}\text{Li}_{2}(z^2),
\eea
it's straightforward show the result is
\bea
I_{\rm AdS,EH}=\frac{1}{8\pi G_N}\[4\h\log\frac{\e}{b}-\im\(\text{Li}_{2}(e^{2\im\h})-\text{Li}_{2}(e^{-2\im\h})\)\].
\eea
 Note that the two edges should be identified once the wedge region is excised. Therefore, when computing the Hayward terms, an overall angle of $\pi$ should be subtracted. A useful convention is dividing it into two identical pieces, as illustrated below. Given that
\be
n_{1}\cdot n_{0}=-\frac{\e}{b}, \quad\sqrt{\g}=\frac{1}{b\ \e}
\ee
the contribution of the corner term is given by
\be
I_{Q_{1,2}\cap \pcM}
=-\frac{2\times 1}{8\pi G_N}\int_{-\h}^{\h}d\q \sqrt{\g}\(\arccos\(n_1\cdot n_0\)-\frac{\pi}{2}\)
=-\frac{\h}{2\pi G_N}.
\ee
which is of constant order. Although we include this term here for completeness, the contribution of this term will be skipped, as  we are primarily concerned with the leading-order logarithmic divergence arising from the Weyl anomaly of the dual BCFT.  When the deficit  angle $\eta$ is small, the leading on-shell action is given by
\bea
I_{gravity, W}&=&I_{\rm AdS,EH}+I_{Q_{1,2}\cap \pcM}\nn
&=&\frac{1}{8\pi G_N}\[4\h\log\frac{\e}{b}-\im\(\text{Li}_{2}(e^{2\im\h})-\text{Li}_{2}(e^{-2\im\h})\)-4\h\]\nn
&\approx&\frac{\h}{4\pi G_N}\log\frac{\e^2}{4\csch^2\b}+\mathcal{O}(\h^3).
\eea

\section{The detail of the integral}
\label{append:B}
\begin{figure}[htbp]
\centering
\subfigure[$\left|z\right|<\left|z_i\right|$]
{
\begin{minipage}[b]{0.4\textwidth}
\includegraphics[width=1\textwidth]{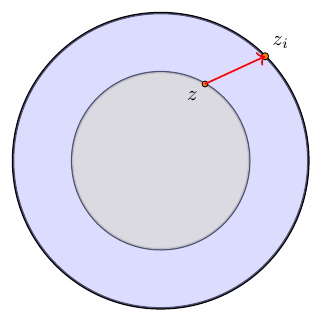}
\end{minipage}\label{figures_01}
}
\subfigure[$\left|z\right|>\left|z_i\right|$]
{
\begin{minipage}[b]{0.4\textwidth}
\includegraphics[width=1\textwidth]{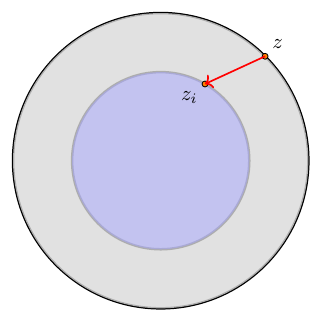}
\end{minipage}\label{figure_02}
}
\caption{Two cases of the integral}
\label{figure_integral}
\end{figure}

\renewcommand{\theequation}{B.\arabic{equation}}
\setcounter{equation}{0}
In this appendix we will show the detail of the following integral 
\bea
\text{I}_{\rm Euc}=\int_{\pcM}\frac{\sqrt{g^{(0)}}d^{2}z}{\left|z-z_{1}\right|^2\left|z-z_{2}\right|^2},
\eea
which is defined on the whole complex plane with the metric
\bea
ds^2_{bdy}=g^{(0)}_{ij}dx^{i}dx^{j}=dz d\bar{z}.
\eea
This integral can also be evaluated using the Feynman parameter technique \cite{He:2019vzf}, with further details about this integral provided in \cite{Chen:2018eqk}. Without loss of generality, we assume  
\be
\left|z_2\right|\geq \left|z_1\right|,
\ee
and introduce  the following polar coordinate
\bea
z=\r e^{\im\q},\quad \bar{z}=\r e^{-\im\q}.
\eea
Using polar coordinates, the integral can be evaluated by first completing the integration over $\theta$. The resulting primitive function is given by
\bea
f(\r,\q)&=&f_{1}(\q,\r)+f_{2}\(\q,\r\)+f_{3}\(\q,\r\)\\
f_{1}(\q,\r)&=&\frac{\q\(\left|z_1\right|^2\left|z_2\right|^2-\r^4\)}{\(\left|z_1\right|^2-\r^2\)\(\bar{z}_1 z_2-\r^2\)\(z_1\bar{z}_2-\r^2\)\(\left|z_2\right|^2-\r^2\)}\nonumber\\
f_{2}(\q,\r)&=&\frac{\im}{\left|z_1\right|^2-\r^2}\[\frac{\bar{z}_1}{\bar{z}_{12}\(\bar{z}_{1}z_2-\r^2\)}\log\(\bar{z}_1-\r e^{-\im \q}\)-\frac{z_1}{z_{12}\(z_{1}\bar{z}_2-\r^2\)}\log\(z_1-\r e^{\im \q}\)\]\nonumber\\
f_{3}(\q,\r)&=&\frac{\im}{\left|z_2\right|^2-\r^2}\[-\frac{\bar{z}_2}{\bar{z}_{12}\(z_1\bar{z}_{2}-\r^2\)}\log\(\bar{z}_2-\r e^{-\im \q}\)+\frac{z_2}{z_{12}\(\bar{z}_1 z_{2}-\r^2\)}\log\(z_2-\r e^{\im \q}\)\].\nonumber
\eea
By applying the residue theorem, it's straightforward to show the following: when $0<\r<\left|z_{1,2}\right|$, the second and third term vanish; when $\r>\left|z_{1,2}\right|$, the logarithm function in above contributes $\pm 2\pi\im$, as illustrated in the Figure \ref{figure_integral}. the contribution of the integral consists of three parts, depending on whether the radius $\r$ lies within the specified interval
$ \left( 0, \left|z_1\right| \right) $, 
$ \left( \left|z_1\right|, \left|z_2\right| \right) $ 
or $ \left( \left|z_2\right|, \infty \right) $. The precise result is given by
\bea
\text{I}_{\rm Euc}&=&\text{I}^{1}_{\rm Euc}+\text{I}^{2}_{\rm Euc}+\text{I}^{3}_{\rm Euc}\\
\text{I}^1_{\rm Euc}&=&\pi \int_{0}^{\(\left|z_1\right|-\delta\)^2}\frac{\(\left|z_1\right|^2\left|z_2\right|^2-\r^4\)d\r^2}{\(\left|z_1\right|^2-\r^2\)\(\bar{z}_1 z_2-\r^2\)\(z_1\bar{z}_2-\r^2\)\(\left|z_2\right|^2-\r^2\)}\nonumber\\
\text{I}^2_{\rm Euc}&=&\pi \int_{\( \left|z_1\right|+\delta\)^2}^{\(\left|z_{2}\right|-\delta\)^2}\frac{\(\left|z_2\right|^2-\left|z_1\right|^2\)d\r^2}{\left|z_{12}\right|^2\(\r^2-\left|z_1\right|^2\)\(\left|z_2\right|^2-\r^2\)}\nonumber\\
\text{I}^{3}_{\rm Euc}&=&\pi \int^{\infty}_{\( \left|z_{2}\right|+\delta \)^2}\frac{\(\r^4-\left|z_1\right|^2\left|z_2\right|^2\)d\r^2}{\(\r^2-\left|z_1\right|^2\)\(\r^2-\bar{z}_1 z_2\)\(\r^2-z_1\bar{z}_2\)\(\r^2-\left|z_2\right|^2\)}\nonumber
\eea
where the radial regulator $\delta$ is introduced. Performing the integral over $\r$ and combining all these three terms give the leading result
\bea
\text{I}_{\rm Euc}&=& \frac{2\pi}{\left|z_{12}\right|^2} \log\frac{\left|z_{12}\right|^2}{\(2\delta\)^2}.
\eea
Up to a rescale of the regulator $\e=2\delta$, the result of this integral is
\bea
\text{I}_{\rm Euc}&=& \frac{2\pi}{\left|z_{12}\right|^2} \log\frac{\left|z_{12}\right|^2}{\e^2}.
\eea
When we evaluated the above integral, we have implicitly assumed that the integration is carried out over the entire complex plane. However, if the integration region is restricted to a annulus, $\rho_a\leq \rho\leq \rho_{b}$, the following boundary term should be included: 
\bea
\frac{\pi}{\left|z_{12}\right|^2}\[\log\frac{\(\left|z_{1}\right|^2-\rho_a^2\)\(\left|z_{2}\right|^2-\rho_a^2\)}{\(\bar{z}_{1}z_{2}-\rho_a^2\)\(z_{1}\bar{z}_{2}-\rho_a^2\)}+\log\frac{\(\rho_{b}^2-\left|z_{1}\right|^2\)\(\rho_{b}^2-\left|z_{2}\right|^2\)}{\(\rho_{b}^2-\bar{z}_{1}z_{2}\)\(\rho_{b}^2-z_{1}\bar{z}_{2}\)}\]
\eea
which is of constant order. When $\rho_a\ll\left|z_{1,2}\right|\ll \rho_{b}$, this boundary term can be neglected.

\section{Higher-order corrections}
\label{append:C}
\renewcommand{\theequation}{C.\arabic{equation}}
\setcounter{equation}{0}
In this appendix, we include the higher-order corrections to the on-shell action of both the global wedge and the BTZ wedge. Since the integrand is quite complicated, we will only present the result and compare it with the numerical result.

\subsection*{Global}
In global AdS, we first recall the leading-order result for convenience, 
\bea
-8\pi G_N I_{gravity, W}=-2\h\log\frac{\e^2}{4\sin\frac{w_{12}}{2}\sin\frac{\bar{w}_{12}}{2}}-2\h\(1+\cosh\b\log\(\coth\frac{\b}{2}\)\).
\eea
The sub-leading correction is
\bea
\frac{\h^3}{9}\[5-3\cosh(2\b)-6\cosh(\b)\sinh^2(\b)\log\(\tanh\frac{\b}{2}\)\],
\eea
and the second and third-order corrections are given by
\bea
&&\frac{\h^5 \[23-60\cosh(2\b)+45\cosh(4\b)+30\sinh^2(\b)\(\cosh\b+3\cosh(3\b)\)\log\(\tanh\frac{\b}{2}\)\]}{900},\hspace{2em}\\
&&\frac{\h^7\[32-21\(6+90\cosh(4\b)+\(2\cosh\b-15\cosh(3\b)+45\cosh(5\b)\)\log\(\tanh\frac{\b}{2}\)\)\sinh^2\b\]}{52920}.\hspace{2.5em}
\eea
\begin{table}[htbp]\label{table1}
	\centering
	\tabcolsep=0.35cm
	\renewcommand\arraystretch{1.5}
	\begin{tabular}{|c|c|c|c|c|c|c|}
		\hline
		Set up  & Leading & Sub-leading  & Second  & Third &Numerical\\ \hline
		$\beta=0.01,\eta=1.57$  & 100.257160 & 101.118245  & 101.202616 & 101.216902 &	101.220328\\ \hline
		$\beta=2,\eta=1.57$  & 76.666174 & 78.361057  & 78.521957 & 78.547600 &	78.553275\\ \hline
		$\beta=5,\eta=1.57$  & 57.787430 & 59.507320  & 59.676879 & 59.703443 &	59.712172\\ \hline
	\end{tabular}
	\caption{Comparison between numerical and perturbative result when $\e=10^{-6}$.}
	\label{table:gAdS3num}
\end{table}
As shown in the Table \ref{table:gAdS3num},  once the higher-order corrections are included, the deviation approaches zero. 

\subsection*{BTZ}
In the BTZ black hole, when $\mathcal{G}<1$, the perturbation result at leading order is given by
\bea
-8\pi G_N I_{gravity, W}=-2\h\(\log\frac{\e^2}{4z_H^2\sech^2\b}+1+\cos^{-1}(\tanh\b)\sinh\b\).
\eea
The sub-leading correction is given by
\bea
\frac{\eta ^3 \left(3 \cosh (2 \beta )-6 \sinh (\beta ) \cosh ^2(\beta ) \cos ^{-1}(\tanh (\beta ))+5\right)}{9},
\eea
and the second and third-order corrections are
\bea
&&\frac{\eta ^5 \left(60 \cosh (2 \beta )+45 \cosh (4 \beta )+30 (\sinh (\beta )-3 \sinh (3 \beta )) \cosh ^2(\beta ) \cos ^{-1}(\tanh (\beta ))+23\right)}{900},\hspace{2.5em}\\
&&\frac{\eta^7(1071\cosh(2\beta)+1890\cosh(4\beta)+945\cosh(6\beta)+190)}{105840}\nn
&&-\frac{\eta^7(2\sinh(\beta)+15 (\sinh(3\beta)+3\sinh (5 \beta)))\cosh^2(\beta)\cos^{-1}(\tanh\beta)}{2520}
\eea
\begin{table}[t]\label{table2}
	\centering
	\tabcolsep=0.35cm
	\renewcommand\arraystretch{1.5}
	\begin{tabular}{|c|c|c|c|c|c|c|}
		\hline
		Set up  & Leading & Sub-leading  & Second  & Third &Numerical\\ \hline
		$\beta=0.01,\eta=1.404717$  & 78.668659 & 81.103846  & 81.859245 & 82.259073 &	83.090989\\ \hline
		$\beta=0.1,\eta=1.323866$ &73.765806 & 75.613561 & 76.050023 &
		76.230839 &	76.414456\\ \hline
		$\beta=0.5,\eta=0.981374$  & 53.404152 & 53.960138 & 54.001459 &54.007191 &	54.008366\\ \hline
		$\beta=1,\eta=0.634524$  & 33.402881 & 33.529183 & 33.531837 &
		33.531939 &	33.531944\\ \hline
	\end{tabular}
	\caption{Comparison between numerical and perturbative result in BTZ black hole when $\mathcal{G}<1$ and the parameters are set as $\e=10^{-6},z_H=1$.}
	\label{table:btznum}
\end{table}
 As illustrated in Table \ref{table:btznum}, for a fixed value of $\mathcal{G}$, as the boost parameter $\b$ increases and the range of deficit angle $\h$ decreases, the deviation between the numerical and perturbation results becomes negligible.
 
\newpage

\mciteSetMidEndSepPunct{}{\ifmciteBstWouldAddEndPunct.\else\fi}{\relax}
\bibliographystyle{utphys}
\bibliography{ref.bib}{}
		
\end{document}